\documentclass[]{aa}  

\usepackage{graphicx}
\usepackage{txfonts}
\usepackage{booktabs}

\usepackage[]{hyperref}
\usepackage{placeins}

\begin{document} 

    \title{A prolific solar flare factory}
    \subtitle{Nearly continuous monitoring of an active region nest with Solar Orbiter}

   \titlerunning{A prolific solar flare factory}
   \authorrunning{Finley et al.}

   \author{A. J. Finley\inst{1}
          \and
          A. S. Brun\inst{1}
          \and
          A. Strugarek\inst{1}
          \and
          B. Perri\inst{1}
          }

    \institute{Universit\'e Paris-Saclay, Universit\'e Paris Cit\'e, CEA, CNRS, AIM, 91191, Gif-sur-Yvette, France\\ \email{adam.finley@cea.fr} }

   \date{Received Mar 30, 2025; accepted -- --, 202-}

\abstract{The properties of active region nests, which are locations on the Sun with recurring flux emergence, are poorly constrained by observations from Earth alone. ESA's Solar Orbiter now monitors the far side of the Sun for extended periods of time.  This facilitates observations of the entire Sun.}
{We combined observations from near-Earth satellites and Solar Orbiter to evaluate the contribution of a long-lived active region nest to the global flaring activity of the Sun.}
{We identified a location in Carrington coordinates with episodic bursts of flux emergence throughout 2022. The combined observations allowed a nearly continuous monitoring of this region from April to October, that is, during its most active period. GOES and Solar Orbiter/STIX were used to compare its flaring activity to that of the entire Sun. The region morphology was extracted from SDO/AIA and Solar Orbiter/EUI extreme-ultraviolet images and was combined with magnetic field measurements from SDO/HMI and Solar Orbiter/PHI to assess its unsigned magnetic flux.}
{The active region nest grew in complexity from January to May due to repeated flux emergence events. The peak unsigned magnetic flux was $5\times10^{22}$Mx. The region caused 40\% of the observed solar flares in 2022, including five months in which it produced 50-70\% of all flares over the entire Sun (in the nearly continuous monitoring window). Of the 17 complex flaring NOAA active regions in 2022, this region contained 10, but they occupied less than 20\% of the area in the active latitudes.}
{Active region nests can maintain a high rate of flaring activity for several solar rotations and are more likely to produce complex active regions that can trigger X-class solar flares. Improving the identification and monitoring of long-lived active region nests would benefit space weather forecasts in the short to medium term. }


   \maketitle
%

\section{Introduction}

\begin{figure*}[h!]
    \centering
    \includegraphics[trim=0cm 0cm 0cm 0cm, clip, width=\textwidth]{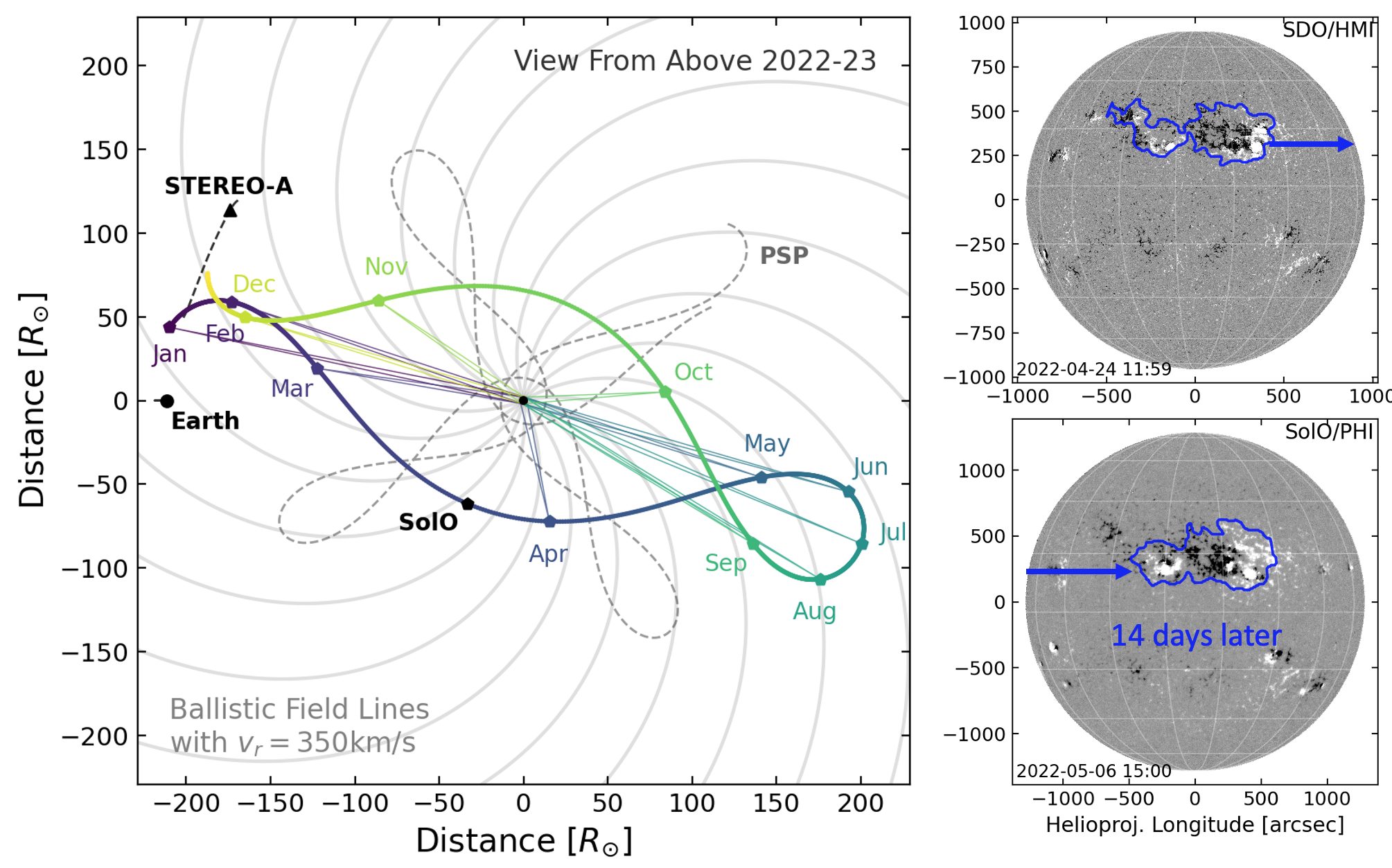}
    \caption{Position of Solar Orbiter in the Earth-Sun reference frame during 2022. From April to October, Solar Orbiter observed the far side of the Sun to Earth. The orbits of Parker Solar Probe and STEREO-A are shown in grey and black, respectively. The right column shows the magnetic field of the AR nest observed from Earth in April and then half a solar rotation later in May as seen by Solar Orbiter.}
    \label{fig:soloOrbit}
\end{figure*}

Active regions (ARs) frequently emerge on the solar surface in close proximity to one another. This nesting affects almost 50$\%$ of all solar ARs \citep{schrijver2008solar}, and it is also commonly observed on other Sun-like stars \citep{berdyugina2005spot, sowmya2021predictions, breton2024tracking}. On the Sun, nested ARs can persist for several solar rotations and have been observed to modulate the solar flaring activity \citep{bai2003hot, gyenge2016active}, the coronal magnetic field topology \citep{benevolenskaya2005formation, tahtinen2024straight, yang2024observing}, and the structure of the solar wind \citep{balthasar1983preferred, neugebauer2000solar}. Whether long-lived nested ARs (hereafter, AR nests) are organised by the underlying dynamo process or form as a stochastic feature of flux emergence remains uncertain \citep[see review of][]{van2015evolution}. 

ARs can cluster because their emergence timescale is shorter (hours to days) than their decay time (weeks to months) and because they have a tendency to emerge in latitudinal bands that migrate towards the equator during the solar cycle \citep{hathaway2011standard}. However, AR nests appear to be more coherent than this: The longitudinal distribution of their flux emergence is organised on timescales of years to decades \citep{berdyugina2003active, mandal2017solar}. The nesting locations have been found to drift in longitude with respect to the Carrington coordinates following the solar latitudinal differential rotation \citep{pojoga2002clustering, zhang2008prediction}. The coherence and longer lifetimes of AR nests were attributed to non-axisymmetries in the generation and storage of the solar dynamo magnetic field \citep{nelson2012magnetic, dikpati2021deciphering, strugarek2023dynamics}. This might be related to the non-linear interactions of Rossby waves \citep{mcintosh2017detection, dikpati2020space, raphaldini2023information} or to the interaction of buoyant magnetic flux with convective motions \citep{jouve2018interactions}. However, because it is challenging for magnetic flux to coherently emerge through the upper convection zone \citep{fan2003dynamic, nelson2014buoyant}, AR nests may need to form closer to the solar surface \citep{vasil2024solar}. In this case, failed deep flux-tube emergences might act like sub-surface reservoirs of magnetic flux that produce coherent flux emergence events when they are perturbed \citep[see][for flux emergence using a dynamo-driven magnetic field]{chen2017emergence}. 

Different methods have been used to identify and classify AR nests in the literature \citep[historically with sunspots, as in][]{castenmiller1986sunspot, brouwer1990sunspot, pelt2005century}. Most previous studies establish a duration for magnetic activity to persist in a given location that is longer than the decay timescale of an isolated emergence; six or more solar rotations were typically used. We define an AR nest as a region of the solar surface with repeated flux emergence that sustains coherent activity over several solar rotations. Consequently, AR nests contain a varying number of individual ARs \citep{bumba1969solar, bumba2000longitudinal}, each with their own degree of magnetic complexity. ARs are given Hale classifications \citep{hale1919magnetic, kunzel1965klassifikation} based on the configuration of the photospheric magnetic field. The label $\alpha$ is given to single-polarity sunspots or groups, and $\beta$ is assigned to regions that contain two or more sunspots with different polarities. These classification are appended by $\gamma$ to signify that the AR contains complex intermixed-polarity flux, and $\delta$ to signify that one of the sunspots in the region contains opposite-polarity flux inside of a shared penumbra. \citet{sammis2000dependence} identified that $\beta\gamma\delta$ ARs more commonly cause the largest solar flares. As the occurrence of complex ARs ($\beta\gamma$, $\beta\delta$, and $\beta\gamma\delta$) peaks slightly later in the solar cycle than simple ARs ($\alpha$ and $\beta$), \citet{jaeggli2016magnetic} inferred that repeated emergences may be required to produce complex regions. In this case, AR nests could be a source of magnetic complexity and consequently produce stronger and more frequent flares than isolated ARs.

As the Sun rotates every 27.28 days when viewed from Earth, ARs are hidden from view for about two weeks. Helioseismology provides an indirect measurement of far side activity \citep[e.g.,][]{lindsey2017seismic, yang2023imaging}, but lacks the sensitivity that is required to monitor AR evolution. This limits our ability to study the long-term behaviour and characteristics of AR nests. With the launch of ESA's Solar Orbiter in 2020 \citep{muller2020solar}, remote-sensing observations became periodically available for the far side of the Sun (see Fig. \ref{fig:soloOrbit} and \citealp{zouganelis2020solar}). Whilst on the far side, Solar Orbiter takes full-disk images at extreme-ultraviolet (EUV) wavelengths with the Extreme Ultraviolet Imager (EUI) \citep{rochus2020solar} and measures the photospheric magnetic field with the Polarimetric and Helioseismic Imager (PHI) \citep{solanki2020polarimetric}. The solar x-ray flux is monitored with the Spectrometer Telescope for Imaging X-rays (STIX) \citep{krucker2020spectrometer}, which can characterise and locate flare sources on the Sun. Observations from Solar Orbiter compliment those available from space-based telescopes around Earth, such as the Atmospheric Imaging Assembly (AIA) \citep{lemen2012atmospheric} and the Helioseismic and Magnetic Imager (HMI) \citep{scherrer2012helioseismic} on board Solar Dynamics Observatory (SDO), as well as the X-ray sensors on board the Geostationary Operational Environmental Satellites (GOES). There are also observations away from the Sun-Earth line from the Extreme Ultraviolet Imager (EUVI) on board the Solar TErrestrial RElations Observatory (STEREO-A) \citep{kaiser2008stereo}. By combining observations from these instruments, we can build a complete picture of the magnetic activity over the entire solar surface. 

\begin{figure*}
    \centering
    \includegraphics[trim=0cm 0cm 0cm 0cm, clip, width=\textwidth]{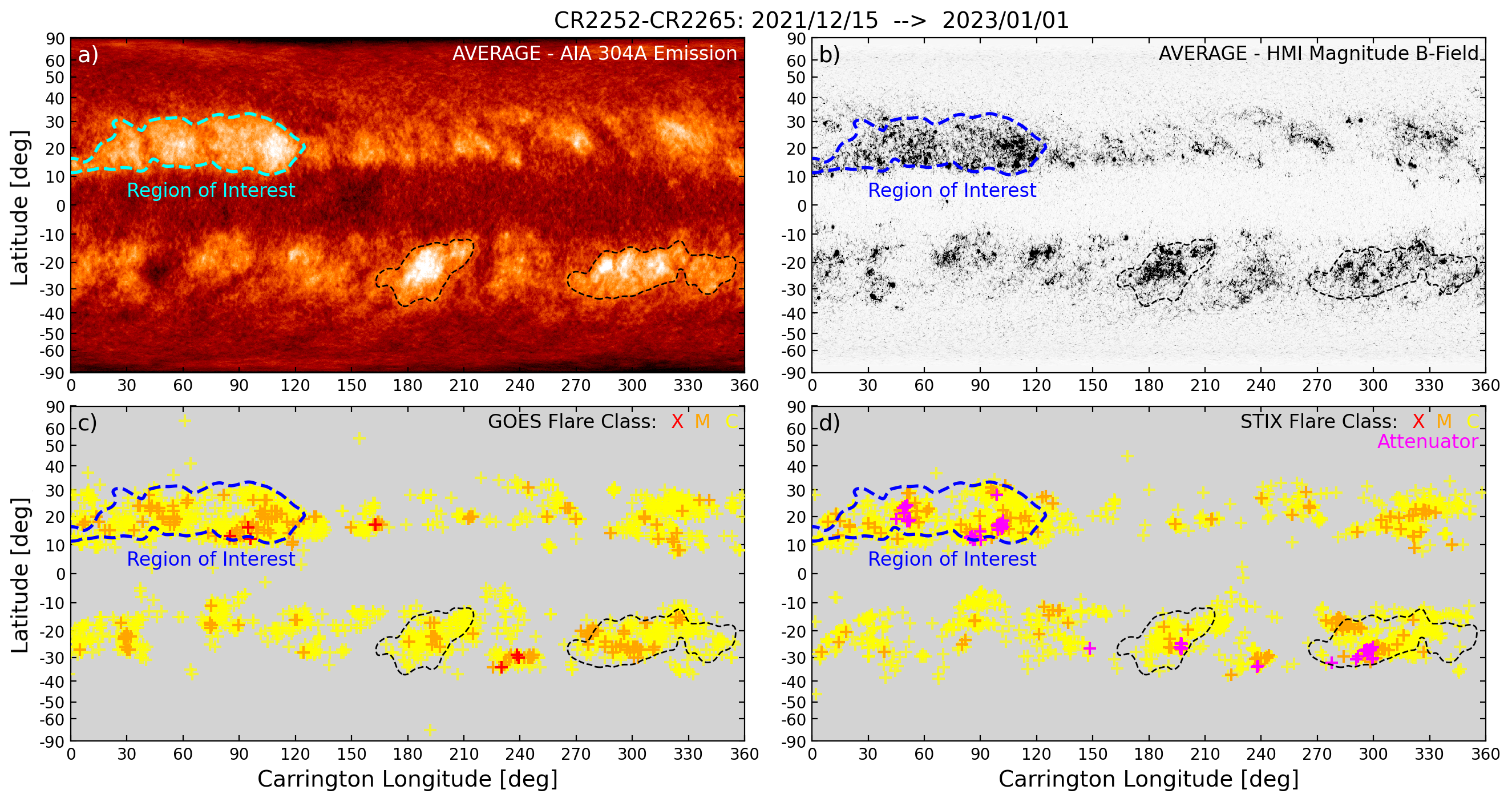}
    \caption{Distribution of magnetic activity in Carrington coordinates in 2022 (sine latitude grid). Panels (a) and (b): Averaged 304$\AA$ EUV emission and magnetic field magnitude, respectively. The average EUV emission is used to identify hot spots that might be related to AR nests. They are highlighted with a dashed black contour. The region of interest for this study is highlighted with a thicker cyan (or blue) contour. Panels (c) and (d): Comparison of the distribution of solar flares observed by GOES and STIX.}
    \label{fig:average_sum}
\end{figure*}

We used observations from Solar Orbiter and near-Earth satellites to characterise the flaring, morphology, and magnetic field evolution of an AR nest in 2022. Figure \ref{fig:soloOrbit} shows the position of Solar Orbiter relative to the Earth throughout 2022. From April to October, Solar Orbiter acted as a far side monitor to Earth, which allowed the AR nest to be near-continuously monitored as it rotated between the two fields of view; an example is shown in Fig. \ref{fig:soloOrbit}. In Section 2 we detail the method used to identify the AR nest and assemble the EUV observations we used. In Sections 3 and 4 we assess the flaring activity of the AR nest and its magnetic field evolution, including the growth in the magnetic complexity. In Section 5 we compute the fraction of solar flares that originated in the AR nest versus the entire solar surface. We conclude that short- to medium-term space weather forecasts would benefit from the early identification and continued monitoring of similar AR nests.

\section{Identification and time-evolution of the AR nest}\label{sect:identify}

In order to locate candidate AR nests, we devised a simple method using Carrington maps of EUV emission. The maps were produced from AIA full-disk observations using a weight kernel that favoured observations closer to the sub-solar point. Therefore, these maps represent a subset of solar activity viewed from Earth during a Carrington rotation (CR). We averaged the maps from 2022 (CR 2252 to 2265) using the AIA 304$\AA$ wavelength. This revealed hot spots of activity in Carrington coordinates that remained EUV bright for several CRs. Figure \ref{fig:average_sum}, panels a and b, show the averaged 304$\AA$ emission from AIA along with the averaged photospheric magnetic field magnitude from HMI. Hot spots in the EUV emission are highlighted in both panels using contours at the 90th percentile value. These regions correlate with the clustering of the average magnetic field in HMI. Flux emergence can be more precisely located using the magnetic field magnitude, but the hot spots were easier to distinguish in the EUV emission. We focused on the largest of these regions, which is labelled ``region of interest''. This hot spot underwent repeated flux emergence over several solar rotations, as detailed in Section \ref{sect:magflux}, and it thus satisfied our definition of an AR nest. The longitudinal extent of the AR nest identification was constrained by this region of interest. This is not a unique solution, however, and warrants future exploration.

\begin{figure*}
    \centering
    \includegraphics[trim=1cm 1cm 0cm 1cm, clip, width=\textwidth]{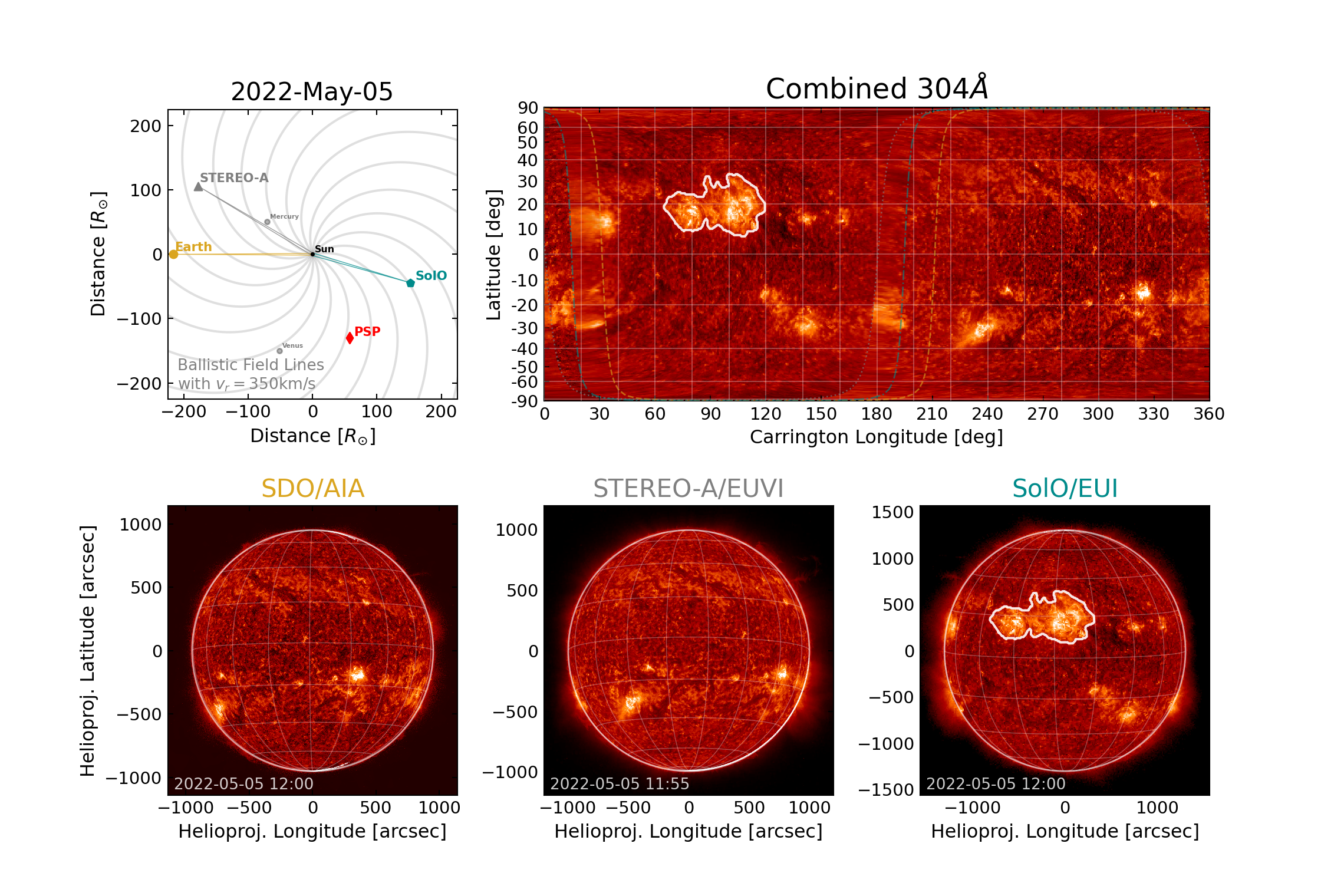}
    \caption{Example of a combined EUV 304$\AA$ observation of the Sun on 5 May 2022, with the AR nest in view of Solar Orbiter on the far side from Earth. The upper left panel contains the orbital configuration of Solar Orbiter, STEREO-A, Parker Solar Probe, and the planets around the Sun, as viewed from above. The three lower panels contain full-disk images from AIA, EUVI, and EUI, and the AR nest is encircled by a white contour. These observations are combined in the Carrington map in the upper right panel.}
    \label{fig:focusEUV}
\end{figure*}

The region of interest is confined to a latitudinal band of about 20$^{\circ}$ in width, but extends over 120$^{\circ}$ in longitude. The longitudinal spread has two contributing factors: 1) multiple flux emergence events that were each slightly separated in longitude, and 2) surface differential rotation advecting the emerged magnetic field with respect to the Carrington coordinates. We followed the morphological evolution of the AR nest by combining daily full-disk images from AIA and EUI. EUI data were scaled to match AIA by normalising overlapping pixels. Figure \ref{fig:focusEUV} shows a combined map of 304$\AA$ observations from 5 May, when the AR nest was in view of Solar Orbiter on the far side of the Sun to Earth. EUVI data were used to fill the gap between the two fields of view, allowing the entire Sun to be viewed at once. The full-disk images were reprojected into Carrington coordinates using the SunPy package \citep{sunpy_community2020} and combined with a weight kernel that favoured observations closer to the sub-solar point for each observer. The shape of the AR nest was extracted by performing a Gaussian smoothing on the combined Carrington map and selecting a brightness contour in the target region. The value of the brightness contour was chosen to outline the majority of the EUV emission in the AR nest and was held at a constant value throughout the study. In some cases, the highlighted area includes some of the surrounding quiet Sun. We extracted the AR nest morphology in this way for every day in 2022 with at least one observation from AIA or EUI. 

\begin{table}
    \caption{NOAA AR numbers and their associated Hale classification(s) for the AR nest from CR 2253 to 2259.}
    \centering
    \begin{tabular}{c  c | c }
    \hline
    CR                  & Start Date    &       NOAA AR                 \\      
    Number              & [d/m/2022]    &       Number(classifications)       \\ 
    \hline
    2253                & 11/Jan        &    12936($\beta$), 12938($\beta$/$\alpha$)  \\ \\
    2254                & 08/Feb        &    12954($\alpha$), 12955($\alpha$),\\
                        &           &    12956($\beta$/$\alpha$)                  \\ \\
    2255                & 07/Mar        &    \textbf{12975}($\beta\gamma\delta$/$\beta\gamma$), \textbf{12976}($\beta\delta$/$\beta$)  \\  \\
    2256                & 03/Apr        &    \textbf{12993}($\beta\gamma\delta$/$\beta\gamma$/$\beta$), \textbf{12994}($\beta\gamma$/$\beta$), \\
                        &           &    12995($\beta$/$\alpha$), 12996($\beta$/$\alpha$)  \\  \\
    2257                & 01/May        &    \textbf{13014}($\beta\gamma\delta$/$\beta\delta$/$\beta\gamma$/$\beta$), 12015($\beta$/$\alpha$)  \\
                        &           &    \textbf{13017}($\beta\gamma$/$\beta$), 13011($\beta$/$\alpha$)  \\  \\
    2258                & 28/May        &    13030($\beta$), \textbf{13032}($\beta\gamma$/$\beta$/$\alpha$),\\
                        &           &    13033($\beta$/$\alpha$), 13034($\beta$/$\alpha$) \\  \\
    2259                & 24/Jun        &    13053($\beta$), 13057($\beta$/$\alpha$)  \\
    \hline
    \end{tabular}
    \tablefoot{Each classification given to an AR while visible to Earth is listed in the form (X/X/X), from most to least complex. Regions with complex classifications, either $\beta\gamma$, $\beta\delta$, or $\beta\gamma\delta$, are highlighted in bold.}
    \label{tab:complexity}
\end{table}

\begin{figure*}
    \centering
    \includegraphics[trim=0cm 0cm 0cm 0cm, clip, width=\textwidth]{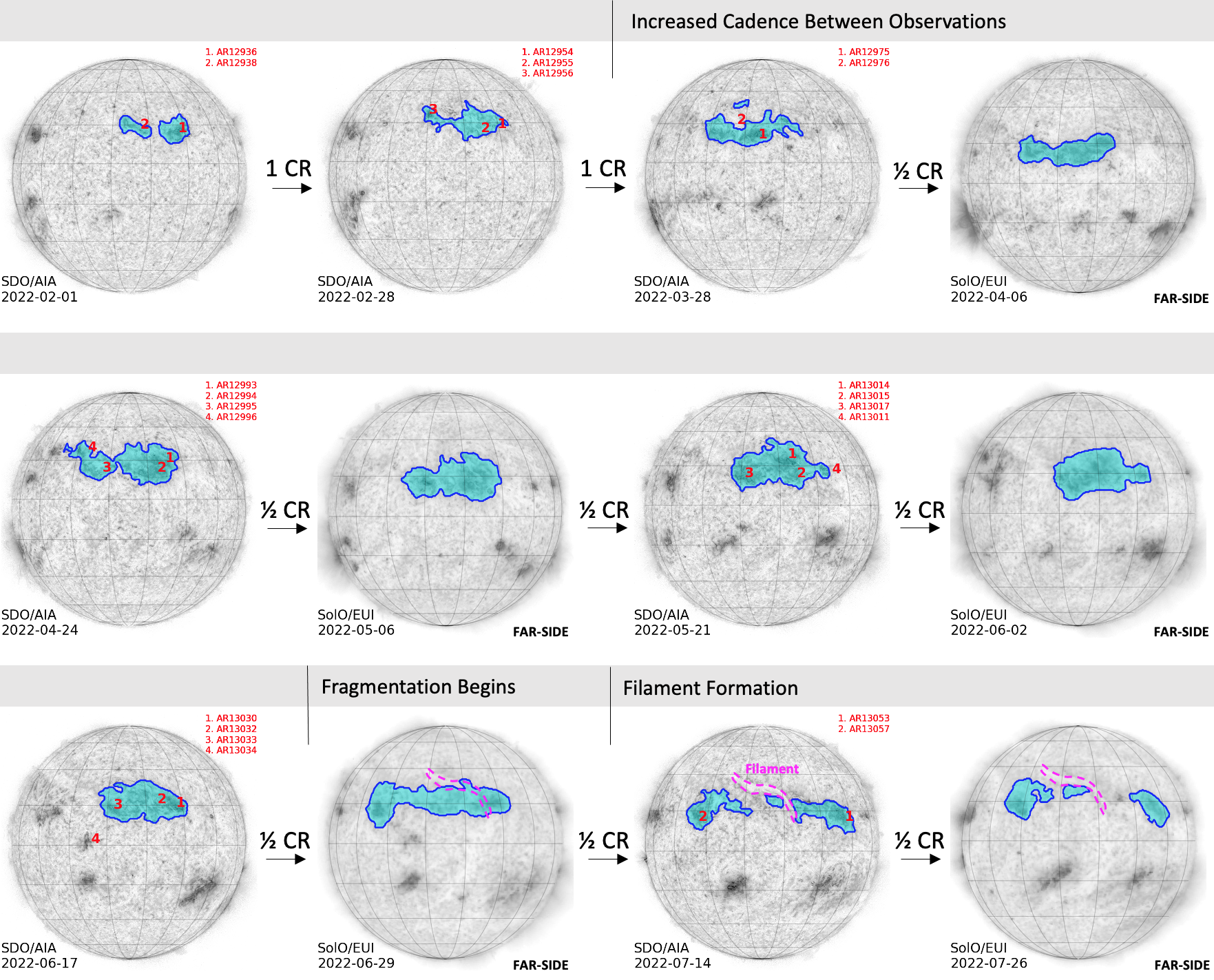}
    \caption{Timeline of EUV 304$\AA$ observations for the AR nest from February to July 2022 (greyscale). From April onwards, the sequence of full-disk images alternates between AIA and EUI (on the far side of the Sun) with roughly two weeks between each image. Each image is centred on 90$^{\circ}$ Carrington longitude with the AR nest highlighted using a filled blue contour. When visible from Earth, the NOAA AR numbers for this region are annotated in red. The latitude of Solar Orbiter varied by a few degrees in this sequence, which changes the apparent latitude of the AR nest.}
    \label{fig:summary}
\end{figure*}

\begin{figure*}
    \centering
    \includegraphics[trim=0cm 0cm 0cm 0cm, clip, width=\textwidth]{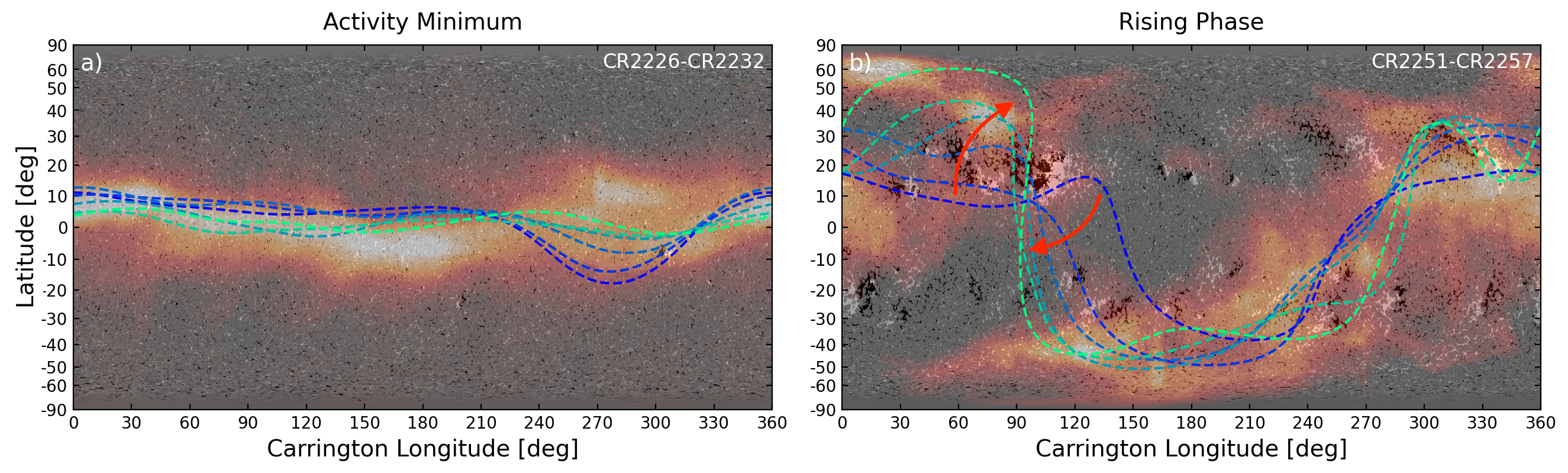}
    \caption{Evolution of the heliospheric current sheet during (a) solar minimum in late 2019 and (b) the rising phase of sunspot cycle 25 in early 2022, when the AR nest emerged. Each panel contains information spanning six CRs and is labelled in the top right corner of each panel. The background greyscale shows a HMI magnetogram from the middle of each CR range. The orange colour gradient shows the average LASCO-C2 scattered white-light brightness at three solar radii during each CR range. The dashed coloured lines show the heliospheric current sheets from potential field source surface models driven by the six corresponding CR magnetograms (blue to green). The impact of the AR nest on the current sheet is highlighted with red arrows.}
    \label{fig:HCS_evol}
\end{figure*}

\begin{figure*}[!h]
    \centering
    \includegraphics[trim=0cm 0cm 0cm 0cm, clip, width=0.48\textwidth]{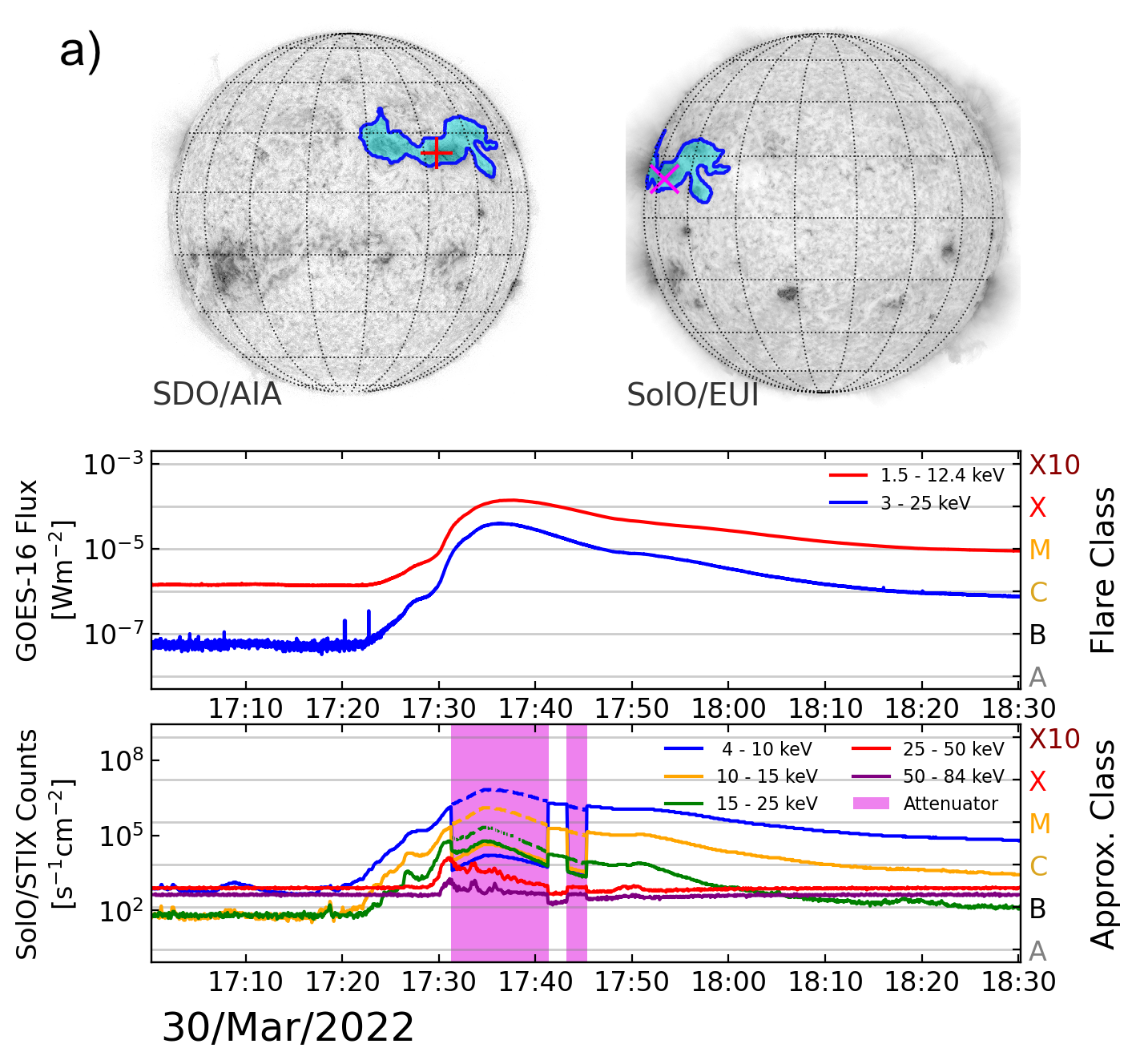}
    \includegraphics[trim=0cm 0cm 0cm 0cm, clip, width=0.48\textwidth]{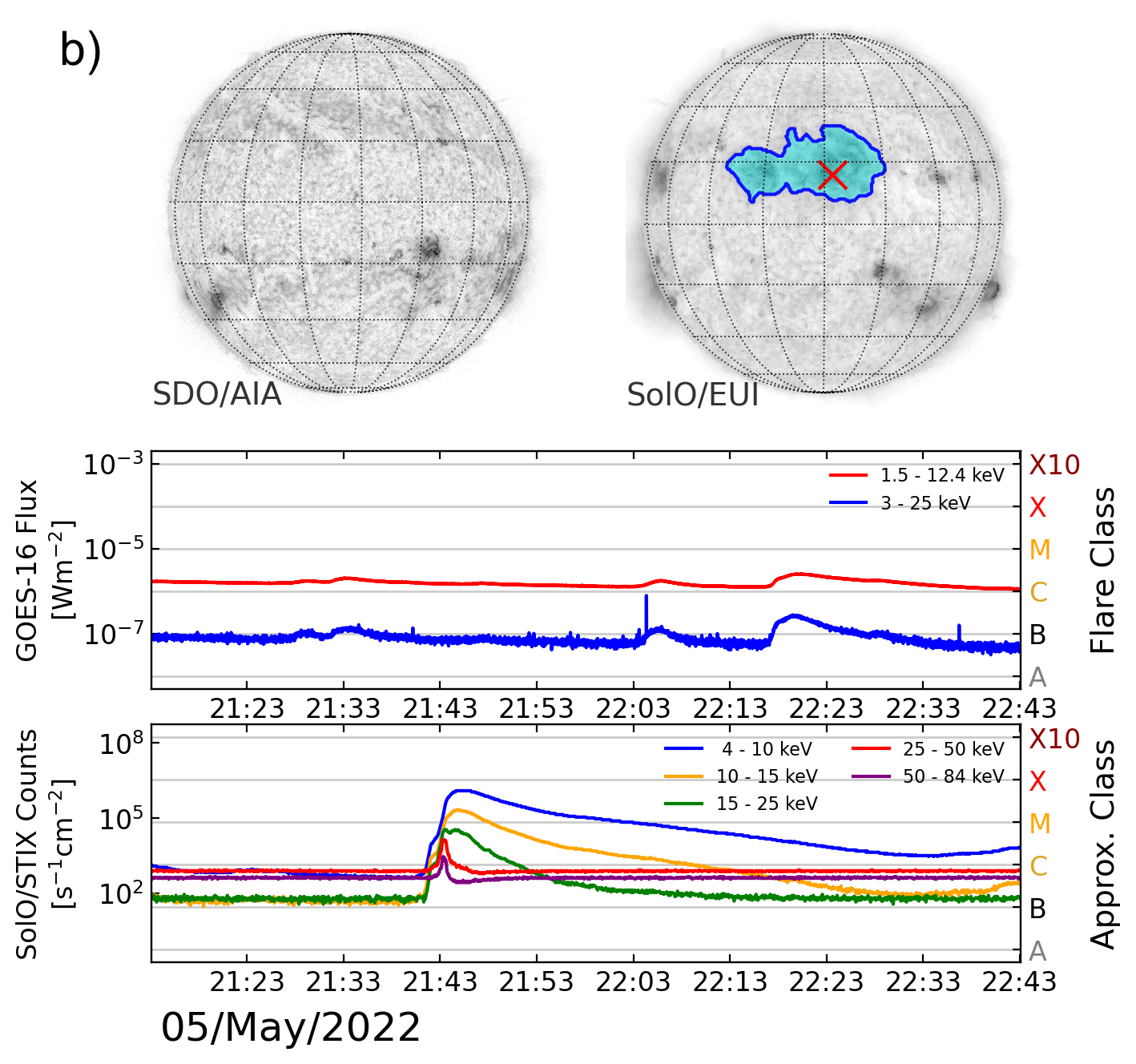}
    \caption{Solar flare light curves and locations as viewed from Earth and Solar Orbiter. The times are given in UTC. Panel (a): X-class flare from 30 March 2022, whilst Solar Orbiter was in quadrature with Earth. The magenta background indicates the time ranges when the attenuator was inserted. The dashed lines show the corrected count rate based on the pre- and post-attenuator values. Panel (b): M-class flare from 5 May 2022, when Solar Orbiter was on the far side of the Sun to Earth. }
    \label{fig:light_curves}
\end{figure*}

Figure \ref{fig:summary} displays a series of full-disk EUV images from AIA and EUI with the AR nest identified using our method. After March, this sequence contains alternating images from AIA and EUI as the AR nest rotated between their respective fields of view; each image is then roughly two weeks apart. When visible to Earth, the ARs in the AR nest were given NOAA AR numbers\footnote{https://www.swpc.noaa.gov/products/solar-region-summary}. These numbers are annotated in Fig. \ref{fig:summary} and are listed in Table \ref{tab:complexity} along with their Hale classifications. This shows a continuity between NOAA ARs numbers that we further discuss in Section \ref{sect:results}. 

Activity began in the AR nest on 23 January when a small pair of ARs (AR12936 and AR12938) rotated into Earth's view. These ARs emerged on the far side of the Sun and were not identified in the available far side helioseismology. Solar Orbiter was positioned between Earth and STEREO-A at this time, and it therefore provided no additional information. After two solar rotations, continued flux emergence had created an elongated patch in EUV that was visible to Solar Orbiter in April as it moved to the far side of the Sun (see Fig. \ref{fig:summary}). From late April until mid-June, the AR nest remained morphologically similar and spanned an area of about $2\%$ of the entire solar surface (the EUV contour area is much larger than the sunspot area, $\sim$0.1\%). After this, the emerged magnetic flux fragmented, leading to the formation of a filament in July. We speculate that the steady separation of the fragmented flux may be due to the AR nest forming from a large concentrated bundle of magnetic field lines beneath the solar surface (a sub-surface flux rope) that then emerged through the photosphere. After the decay of these fragments, the AR nest experienced additional flux emergence in September and December. Appendix \ref{ap:stack_plots} summarises the distribution of magnetic activity in the northern and southern activity bands with stack plots of the daily combined EUV maps in time versus Carrington longitude. Figure \ref{fig:flares_stacked} confirms that this AR nest persisted through 2022 with varying levels of magnetic activity.

The AR nest significantly influenced the evolution of the coronal magnetic field in 2022 (discussed in \citealp{finley2024nested}). To show this, we calculated a series of potential field source surface models \citep{altschuler1969magnetic, schrijver2003photospheric}, with a source surface of 2.5 solar radii, and extracted the shape of the heliospheric current sheet. Figure \ref{fig:HCS_evol}, panel (a), shows a relatively flat heliospheric current sheet from a series of six consecutive CRs that span the recent solar minimum in December 2019. This is reflected in the averaged scattered white-light observations from the Large Angle and Spectrometric Coronagraph (LASCO) on board the Solar and Heliospheric Observatory (SOHO) \citep{domingo1995soho}. Panel (b) shows six CRs during the rising phase in 2022. The emergence of new magnetic flux in the AR nest twists, and subsequently anchors, the heliospheric current sheet above. Both Solar Orbiter and the NASA Parker Solar Probe \citep{fox2016solar} passed through this inclined current sheet during encounter 11 in February 2022 \citep{rivera2024situ}. This is discussed further in Appendix \ref{ap:psp_e11}.

We searched for flux emergence events in the AR nest before and after 2022. This revealed bursts of activity in 2020, 2021, and later in 2023. Carrington maps of EUV emission during these epochs are provided in Fig. \ref{fig:beforeafter}, with the contour from Fig. \ref{fig:average_sum} over-plotted for reference. Following the latitudinal migration of AR emergence over the solar cycle, the flux emergence events in 2020 and 2021 appear at the northern edge of the contour and the those in 2023 at the southern edge. The recurrence of activity in this region over more than three years further supports our classification of this region as an AR nest, and it reinforces the findings from previous studies that investigated the recurring flux emergence over longer timescales \citep[e.g.][]{berdyugina2003active}. However, we limited the duration of this study to 2022 in order to focus on the combined observations with Solar Orbiter.

\section{Solar flare energies and distribution}\label{sect:flares}

\begin{figure*}[!h]
    \centering
    \includegraphics[trim=0cm 0cm 0cm 0cm, clip, width=\textwidth]{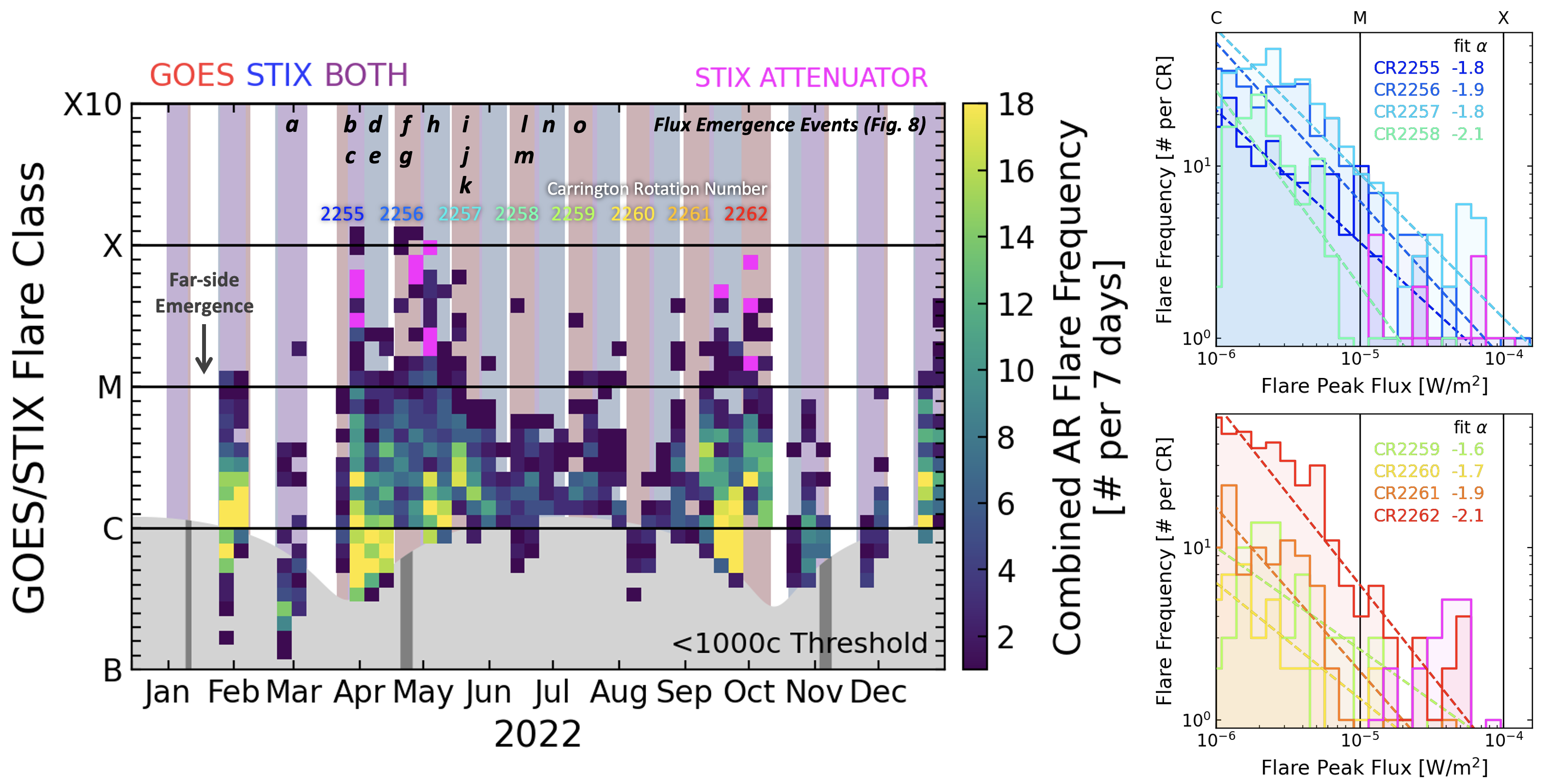}
    \caption{Combined histogram of solar flares observed by GOES and STIX from the AR nest in 2022. Left panel: Visibility to GOES, STIX, or both (red, blue, and purple background colour). Bins that contain one or more flare that triggered the STIX attenuator are highlighted in magenta (the fluxes of these events were not corrected). The STIX 1000 count threshold is shown in light grey. Intervals when STIX was offline are represented by dark grey bars in the count threshold. The flux emergence events from Fig. \ref{fig:mag_summary} are labelled (a) to (o). Right panels: Distribution of solar flares in eight CR bins from 2255 to 2262 that span the period of nearly continuous observations.}
    \label{fig:histogram}
\end{figure*}

The location and intensity of solar flares were taken from the GOES flare list\footnote{Data available: https://www.lmsal.com/solarsoft/latest\_events/} using the Heliophysics Event Knowledgebase \citep{hurlburt2012heliophysics} and the STIX flare list produced by Laura Hayes\footnote{Data available: https://github.com/hayesla/stix\_flarelist\_science/}. Figure \ref{fig:light_curves} shows two flare light curves. Panel (a) contains an X-class flare that was viewed by both observers, and panel (b) contains an M-class flare that was viewed by Solar Orbiter on the far side to Earth. STIX uses an attenuator to reduce the measured intensity during the strongest flares. The X-class flare in Fig. \ref{fig:light_curves}, panel (a), is an example of one such flare that triggered the attenuator on STIX. As there is no standardised method to correct for the count rate when the attenuator is inserted, this was postponed to future work. Attenuated flares are clearly identified in all figures and tables in this study.

STIX reports x-ray counts rather than a flux, like GOES. The 4 - 10 keV count rate measured by STIX was converted into an approximate 1.5 - 12.4 keV GOES x-ray flux using the scaling
\begin{equation}
    \log_{10}(F_{GOES}) = a + b\cdot\log_{10}(N_{STIX} \cdot R^2),
    \label{eq:flare}
\end{equation}
where $F_{GOES}$ is the peak GOES flux in Wm$^{-2}$, $N_{STIX}$ is the peak STIX count rate, and $R$ is the radial distance of Solar Orbiter in au. The parameters $a=-7.7$ and $b=0.59$ were fitted to a collection of flares observed by both GOES and STIX at the start and end of 2022 (when Solar Orbiter was close to Earth). \citet{xiao2023data} fitted similar values of $a=-7.3$ and $b=0.62$ \citep[see also][]{stiefel2025using}. Figure \ref{fig:average_sum}, panels (c) and (d), shows the distribution of solar flares in 2022 observed by GOES and STIX, coloured by flare classification X to C. Their sources are similar, but from April to October, the two instruments observed opposite sides of the Sun. The AR nest and the other EUV hot spots clearly shape the spatial distribution of the solar flaring activity. However, some large M- and X-class flares lie outside of the EUV hot spots (contours). These flares either belong to strong isolated emergences, newly forming hot spots that have yet to influence the averaged EUV emission, or ARs that were missed because the Carrington maps favour observations near the central meridian. 

The GOES and STIX flare lists were combined into weekly distribution functions. Figure \ref{fig:histogram} shows the time-varying distribution of solar flares from the AR nest using the area defined by the region of interest in Fig. \ref{fig:average_sum}. From April to October (CR 2255 to 2262), the AR nest was almost always visible by one or more of the observers, and nearly all of the flaring events were thus captured. When we generated the STIX flare list, a minimum count threshold of 1000 in the 10-15 keV channel was used. Following Eq. (\ref{eq:flare}), this corresponds to a time-varying x-ray flux threshold because the distance of STIX varies throughout the year. At the farthest distance of Solar Orbiter from the Sun, this sets a minimum threshold for completeness at the C1 GOES flux. As we focus on C-, M-, and X-class flares, this has no impact on our conclusions.

We captured two periods of high activity from the AR nest during the nearly continuous coverage from April to October: one burst at the start of this window, and the other at the end (see Fig. \ref{fig:histogram}). During these periods, GOES measured tens of large M-class flares and a few X-class flares. STIX observed a similar distribution with several flares causing the attenuator to be inserted; they are likely strong M- and X-class events. The first period of high activity, starting in March, was the most prolific. The frequency of C- and M-class flares remained high until May. An X-class flare was observed on 30 March from AR12975 ($\beta\gamma\delta$ classification). This flare was recorded by both GOES and STIX (with attenuator inserted) while Solar Orbiter was moving to the far side (shown in Fig. \ref{fig:light_curves}, panel a). Two more X-class flares were observed on 17 April (east limb to Earth) and 30 April (west limb) from AR12994 ($\beta\gamma$). These flares occurred as the AR nest rotated between the two fields of view. The next strong (attenuated) flare was observed by STIX on 3 May from AR12994 on the far side of the Sun. Fig. \ref{fig:light_curves}, panel (b), displays a slightly smaller (but unattenuated) M-class flare that occurred two days later on 5 May. The higher-energy channels associated with the impulsive non-thermal x-ray emission peak as the thermal emission builds up. As with most of the far side events, there was no observable feature in the two GOES channels.

In June, between the two periods of high activity, the region fragmented and formed a large filament (highlighted in Fig. \ref{fig:summary}). This filament partially erupted on 15 July whilst in view of Earth. The second period of high activity started in September and continued until the end of October. It coincided with the return of Solar Orbiter from the far side. This period did not contain any X-class events, but a series of strong M-class flares was produced by AR13112 ($\beta\gamma\delta$) whilst on the far side (activating the attenuator of STIX). This series of flares began on 18 September and continued until the AR nest was visible to Earth on 3 October. 

\begin{figure*}
    \centering
    \includegraphics[trim=0cm 0cm 0cm 0cm, clip, width=\textwidth]{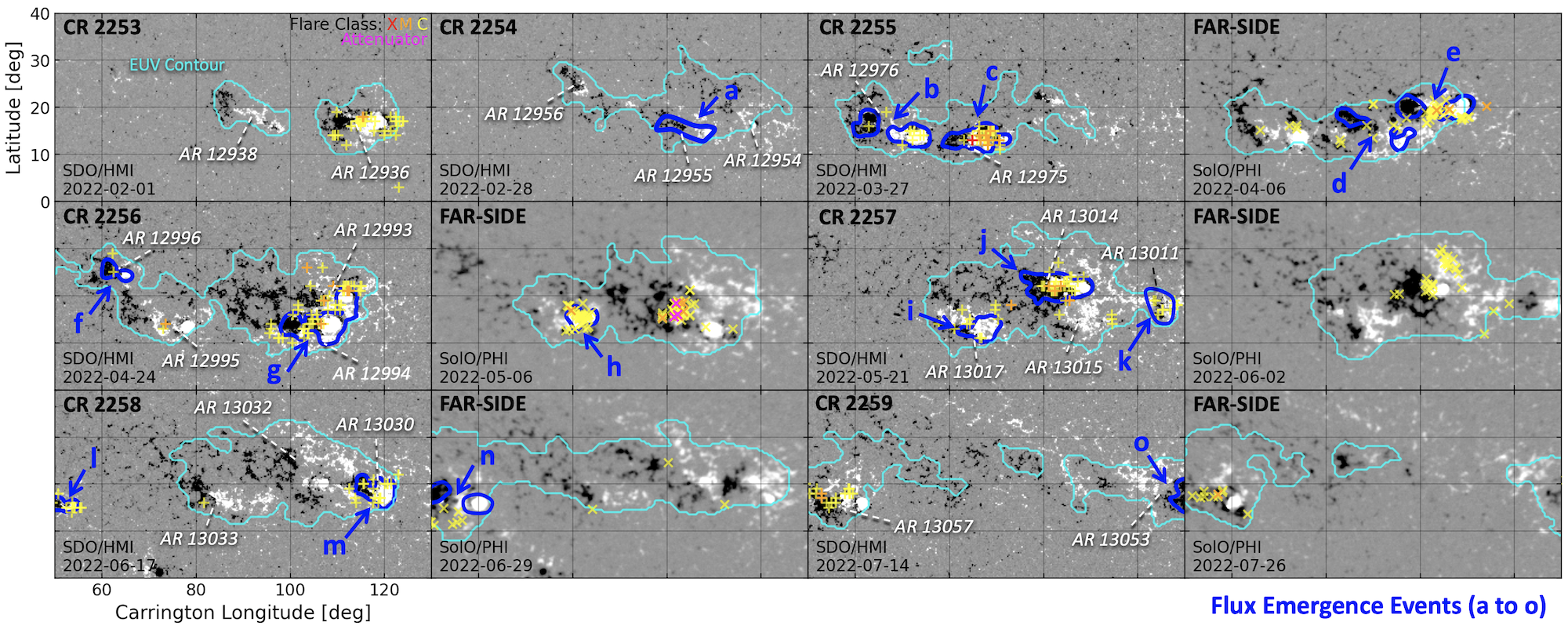}
    \caption{Snapshots of the photospheric magnetic field in the AR nest from HMI and PHI. The snapshots are chosen to match the EUV observations from Fig. \ref{fig:focusEUV} best. The measurements are projected into Carrington coordinates. Solar flares observed by GOES or STIX within three days of each snapshot are shown with crosses that are coloured by flare class. When available, the NOAA AR numbers are annotated. The arrows labelled (a) to (o) highlight flux emergence events between each snapshot.}
    \label{fig:mag_summary}
\end{figure*}

The two right panels of Fig. \ref{fig:histogram} show the peak flux distribution of solar flares from the AR nest during the eight Carrington rotations with nearly continuous observations (CR 2255 to 2262). We fitted the distribution from each CR with a power-law exponent $\alpha$ corresponding to the relation $dN \propto F_X^{\alpha}dF_X$, where $N$ is the flare frequency, and $F_X$ is the flare peak flux \citep[see][and references therein]{aschwanden2012automated}. No attenuated flares from STIX were included in the fitting. The value of $\alpha$ ranges from -1.6 to -2.1 with an average value of $-1.86\pm0.18$, which is consistent with previously derived exponents for soft x-ray flares (e.g. \citealt{aschwanden2016review}). We have less confidence in the values of $\alpha=-1.6,-1.7$ for CR 2259 and 2260 as they were fitted with far fewer flares than the other CRs during the fragmentation and filament formation phase. The value of -2.1 from CR 2262 may be steeper than in reality because high-energy flares are effectively filtered by the STIX attenuator. These flares pile up around the activation energy of the attenuator, and their true peak flux is obscured. 

Outside the period of nearly continuous observations, there were two other notable bursts of activity in the AR nest. The first burst was its initial emergence, with AR12936 and AR12938 ($\beta$ regions) rotating into view of Earth on 23 January. Flaring activity switched from AR12938 to AR12936 on 27 January, perhaps implying that AR12938 was the younger of these two regions. The first flares from the AR nest in 2022 were mostly between GOES C2 and C4 class, and a few reached up to M1. The second notable burst took place in December and continued into 2023. The flare distribution reached as high as M6. AR13169 ($\beta\gamma$) and AR13171 ($\beta$) emerged out of view on the far side of the Sun, but had a clear signature in time-distance helioseismology at 90$^{\circ}$ Carrington longitude. The configuration of the AR nest at the end of 2022 was remarkably similar to its first appearance in January.

\section{Magnetic field evolution and complexity}\label{sect:magflux}

We connected the flaring activity of the AR nest to the evolution of the underlying magnetic field using HMI and PHI observations of the line-of-sight photospheric magnetic field. Carrington maps of the magnetic field were constructed by combining multiple full-disk observations with a weight kernel that favours observations closer to the sub-solar point, that is, where the line-of-sight field is radial. Previous works that combined full-disk measurements between these two telescopes \citep[e.g.][]{loeschl2024first} began to inter-calibrate their magnetic field strengths \citep{vacas2024comparison}. Accordingly, we scaled the full-disk magnetic field observations from PHI by a factor of 1.3 to match the unsigned magnetic flux in the AR nest observed by HMI during the start and end of 2022 (when the AR nest was visible to both observers). The magnetic flux in the AR nest was evaluated each time the region passed the central meridian of one of the observers. From April to October, the combined observations were sufficient to identify the majority of flux emergence events during the first period of high activity (March to June). Figure \ref{fig:mag_summary} shows the photospheric magnetic field of the AR nest during the same time span as the EUV images presented in Fig. \ref{fig:summary}, reprojected into Carrington coordinates. We identified new magnetic flux that appeared in each snapshot by subtracting and smoothing each pair of images in the sequence. Areas with a significant increasing magnetic flux are labelled as flux emergence events with letters (a) to (o). Many of these events are clustered around 100$^{\circ}$ Carrington longitude, with some smaller events towards 70$^{\circ}$ Carrington longitude. This helps us to explain the longitudinal extent and two-lobe morphology of the AR nest in the EUV observations. Most of the large M-class and X-class flares in Fig. \ref{fig:average_sum}, panels (c) and (d), were located near 100$^{\circ}$ Carrington longitude. They correspond to the largest of the flux emergence events. 

\begin{table*}
    \caption{Properties of the AR nest in comparison to the global solar values. Percentage surface area, unsigned magnetic flux, and solar flares are given as the total in the AR nest divided by the global value.}
    \centering
    \begin{tabular}{c | c | c c | c c c c c}
    \hline
    Time Range [2022]                   &       Primary         &       Surface Area                            &       Magnetic Flux                           &       C-class                         &       M-class                         &       X-class &                               Attenuator                              \\      
    Start       -       End [d/m]       &       Observer        &       [x10$^3$Mm$^2$]                         &       [x10$^{22}$Mx]                          &       [\# of events]                              &       [\# of events]                          &       [\# of events]                              &       [\# of events]                          \\      \hline
    1st Period &                &       &       &       &       &       &       \\      
    14/Mar      -       01/Apr  &       Earth   &       76.2    (1.3\%) &       2.8     (7.9\%) &       116     (52.3\%)        &       12      (57.1\%)        &       1       (100.0\%)       &       8       (100.0\%)       \\      
    01/Apr      -       15/Apr  &       Solar Orbiter    &       74.3    (1.2\%) &       4.4     (10.7\%)        &       171     (67.1\%)        &       9       (90\%)  &       0       (-\%)   &       0       (-\%)   \\      
    15/Apr      -       30/Apr  &       Earth   &       117.2   (1.9\%) &       5.1     (16.3\%)        &       137     (66.8\%)        &       26      (57.8\%)        &       1       (50.0\%)        &       0       (-\%)   \\      
    30/Apr      -       14/May  &       Solar Orbiter    &       114.7   (1.9\%) &       4.2     (12.7\%)        &       205     (50.7\%)        &       18      (50\%)  &       1       (33.3\%)        &       14      (93.3\%)        \\      
    14/May      -       27/May  &       Earth   &       117.7   (1.9\%) &       4.5     (13.5\%)        &       160     (69.6\%)        &       8       (80\%)  &       0       (-\%)   &       0       (-\%)   \\      
    27/May      -       10/Jun  &       Solar Orbiter    &       122.4   (2\%)   &       4.2     (13.5\%)        &       63      (67\%)  &       1       (50\%)  &       0       (-\%)   &       0       (-\%)   \\      
    10/Jun      -       23/Jun  &       Earth   &       100.6   (1.7\%) &       3.3     (10.7\%)        &       64      (41\%)  &       2       (66.7\%)        &       0       (-\%)   &       0       (-\%)   \\      
    Fragmentation &             &       &       &       &       &       &       \\      
    23/Jun      -       07/Jul  &       Solar Orbiter    &       152.0   (2.5\%) &       4.1     (13.5\%)        &       33      (42.9\%)        &       0       (0\%)   &       0       (-\%)   &       0       (-\%)   \\      
    07/Jul      -       20/Jul  &       Earth   &       100.6   (1.7\%) &       2.6     (8.4\%) &       52      (24.8\%)        &       4       (40\%)  &       0       (-\%)   &       0       (-\%)   \\              
    20/Jul      -       03/Aug  &       Solar Orbiter    &       93.3    (1.5\%) &       2.4     (8.2\%) &       47      (29.7\%)        &       3       (42.9\%)        &       0       (-\%)   &       0       (-\%)   \\                              
    03/Aug      -       16/Aug  &       Earth   &       56.5    (0.9\%) &       1.2     (4.4\%) &       6       (5.4\%) &       0       (-\%)   &       0       (-\%)   &       0       (-\%)   \\      
    2nd Period &                &       &       &       &       &       &       \\      
    16/Aug      -       30/Aug  &       Solar Orbiter    &       50.8    (0.8\%) &       1.1     (3.8\%) &       50      (11.1\%)        &       3       (5.4\%) &       0       (-\%)   &       0       (-\%)   \\      
    30/Aug      -       13/Sep  &       Earth   &       34.6    (0.6\%) &       1.1     (3.4\%) &       92      (19.6\%)        &       1       (4.2\%) &       0       (-\%)   &       0       (-\%)   \\      
    13/Sep      -       27/Sep  &       Solar Orbiter    &       117.0   (1.9\%) &       4.6     (11.7\%)        &       215     (49.5\%)        &       16      (66.7\%)        &       0       (-\%)   &       6       (54.5\%)        \\      
    27/Sep      -       15/Oct  &       Earth   &       156.9   (2.6\%) &       5.1     (12.5\%)        &       289     (73.4\%)        &       26      (76.5\%)        &       0       (-\%)   &       8       (100.0\%)       \\      
    15/Oct      -       28/Oct  &       Solar Orbiter    &       190.9   (3.1\%) &       4       (8.8\%) &       22      (21\%)  &       0       (-\%)   &       0       (-\%)   &       0       (-\%)   \\      
    \hline
    \end{tabular}
    \label{tab:global_params}
\end{table*}

\begin{figure}[!h]
    \centering
    \includegraphics[trim=0cm 0cm 0cm 0cm, clip, width=0.5\textwidth]{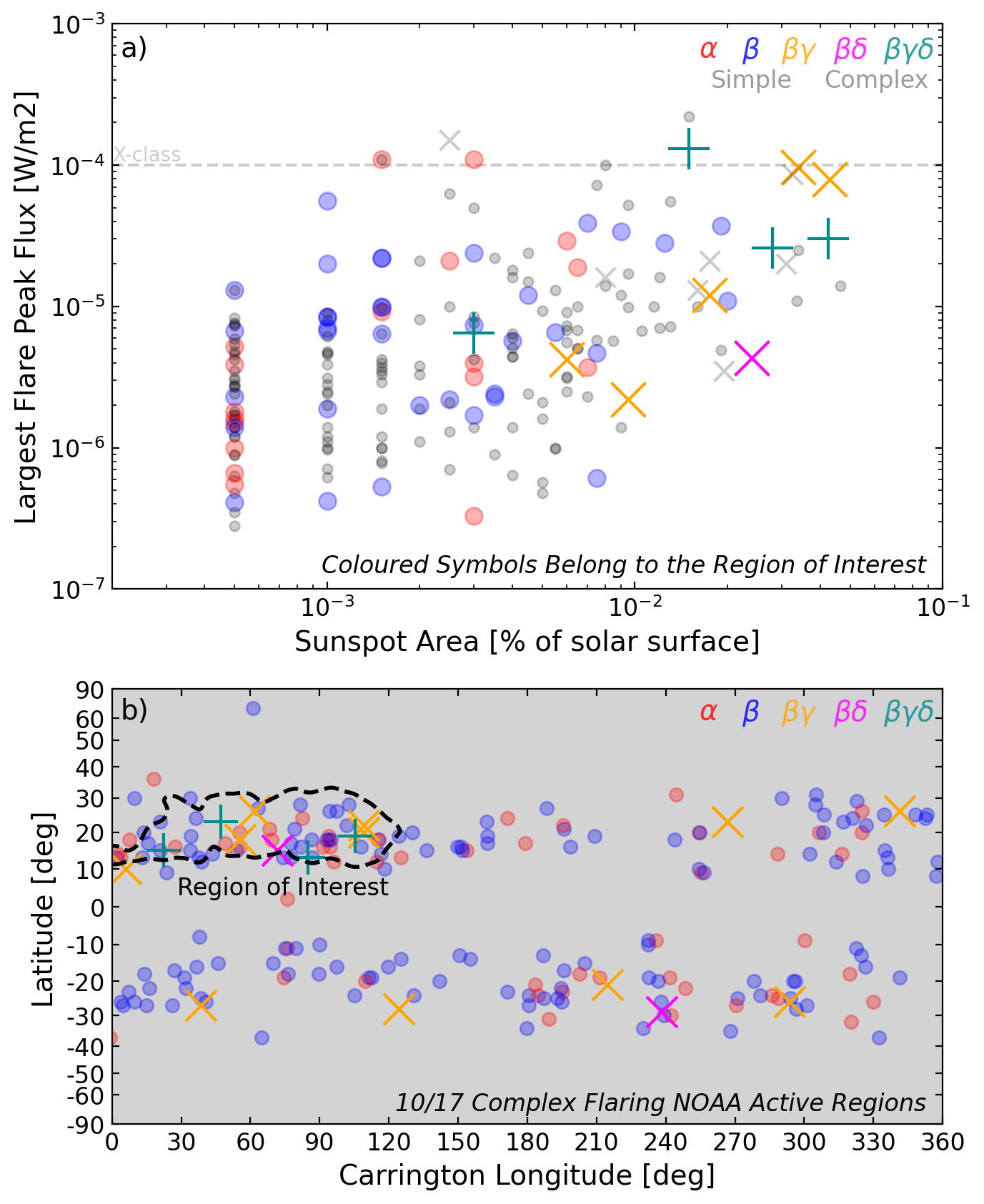}
    \caption{Comparison of the peak flux of the strongest flares from NOAA ARs with their Hale classification. Panel (a): Peak flare flux for each NOAA AR in 2022 vs. their maximum sunspot area. NOAA ARs that were associated with the AR nest are coloured with larger symbols, and the others are shown in grey. Panel (b): Location of these NOAA ARs in Carrington coordinates. }
    \label{fig:complexity}
\end{figure}

The AR nest began in January with the emergence of AR12936 and AR12938 (both $\beta$ regions), and it mostly produced C-class flares. After one solar rotation, these ARs had largely decayed (now AR12954 and AR12956), and flux emergence event (a) had produced AR12955 ($\beta$ that decayed to $\alpha$). This region caused the far side filament eruption on 15 February that was observed at up to six solar radii away by EUI  \citep{mierla2022prominence} along with extended x-rays emission detected in the flux rope by STIX \citep{hayes2024high}. When it was back in view of Earth and Solar Orbiter, only a handful of C-class and one M-class flares where produced from the AR nest (see Fig. \ref{fig:histogram}). At the end of March, the AR nest had a significant burst of flux emergence, labelled (b) and (c) in Fig. \ref{fig:mag_summary}. This formed two large bipoles AR12976 ($\beta\delta$) and AR12975 ($\beta\gamma\delta$), respectively. The complexity of AR12975 might arise from its proximity to previously emerged magnetic flux, whereas AR12976 appeared in an unoccupied area. At this time, the AR nest contained $2.8\times 10^{22}$Mx of unsigned magnetic flux, and it produced 12 M-class flares while visible to Earth (8 flares that triggered the STIX attenuator). The complex AR12975 caused an X-class flare on 30 March. \citet{janvier2023multiple} studied the filament eruption that took place on 2 April from the AR nest (emerging in between AR12975 and AR12976) and took advantage of the multi-viewpoint observations.

In April, nearly continuous observations began with Solar Orbiter on the far side to Earth. Flux emergence events (d) and (e) produced two bipoles, with event (e) in an unoccupied area and (d) emerging in proximity to the decaying AR12975. The flares observed by STIX were mostly concentrated around flux emergence event (e), likely because it contained a stronger concentration of magnetic flux. During this two-week period, the AR nest produced 67\% of all C-class events on the Sun (171 flares) and 90\% of all M-class events (9 flares). The M- and X-class events decreased in number from the previous CR. Returning to Earth's field of view, the AR nest morphology changed drastically due to the decay of ARs, surface motions, and flux emergence events (f) and (g). Event (f) produced AR12996 ($\beta$), which trailed the bulk of the magnetic flux in the AR nest. Event (g) supplied a new bipolar region below the existing ARs and either enhanced the positive polarity of the AR formed by event (e) or caused another smaller bipole, with the negative polarity hidden inside the main bipolar region. This is further discussed in Section \ref{sect:highlight}. This formed AR12993 ($\beta\gamma\delta$) and AR12994 ($\beta\gamma$). X-class flares were produced by AR12994 whilst it rotated into Earth view on 17 April and again on 30 April as it left Earth view (east limb for Solar Orbiter). The AR nest contained $5.1\times 10^{22}$Mx of unsigned magnetic flux (about 16\% of the total surface magnetic flux) and produced the majority of flares ($\sim 50-60\%$) over the entire Sun, but the distribution shifted towards more M-class events (17 more than in the previous two-week period) and fewer C-class events (34 events less).

The following solar rotation in May initially had a reduced flux emergence rate with only event (h) on the far side. This caused the region to diffuse significantly (decreasing to $4.2\times 10^{22}$Mx). A burst in the flux emergence in the following weeks, events (i), (j), and (k), then created AR13017 ($\beta$), AR13014 ($\beta\gamma\delta$), and AR13011 ($\beta$). Despite the complexity of AR13014, the overall flaring rate of the AR nest dropped from over 70 C-class flares per week to about 30 flares after an initial few M-class flares (see Fig. \ref{fig:histogram}). From June to July, the region fragmented with flux emergence events (l), (m), (n), and (o). This split the region into two and formed a filament in between. The unsigned magnetic flux dropped to $1.1\times 10^{22}$Mx after the fragmentation. Activity returned to the AR nest during the second period of high activity in September (climbing back up to $4\times 10^{22}$Mx of unsigned magnetic flux). During this epoch, the flux emergence events were closer to 60$^{\circ}$ Carrington longitude. They might have shifted due to the solar differential rotation. The number of M- and C-class flares was similar overall to the first period of activity.

To compare the magnetic complexity of the ARs that formed in the AR nest to the complexitiy elsewhere on the Sun, we identified the largest flares from each NOAA AR in 2022 and collected their Hale classifications. In Fig. \ref{fig:complexity}, panel (a), we show the distribution of NOAA ARs with their highest flare fluxes scattered against their maximum sunspot area. The simple Hale classifications of $\alpha$ and $\beta$ are identified differently (dots) to the more complex $\beta\gamma$, $\beta\delta$, and $\beta\gamma\delta$ configurations (crosses). There is a trend for larger sunspot areas to produce larger flares and for larger sunspot areas to be more complex \citep[as shown by][]{sammis2000dependence}. Identifying the ARs from the AR nest in colour, we show that the AR nest contained a significant fraction of the large complex regions in 2022. This is apparent in Fig. \ref{fig:complexity}, panel (b), where the complex regions are visibly grouped inside the region of interest. Seven of these complex regions, three $\beta\gamma\delta$, three $\beta\gamma$, and one $\beta\delta$ (see Table \ref{tab:complexity}), formed during the first period of high activity (explored in Fig. \ref{fig:mag_summary}). In total, the AR nest contained 10 of the 17 complex flaring ARs from 2022, which suggests that the repeated emergence of magnetic flux facilitated the creation of more complex regions \citep[as discussed in][]{title2016flare}. This motivated a more detailed analysis of the free magnetic energy built up by this region (Blaise et al. in preparation).

\section{Discussion}\label{sect:results}

\subsection{Flaring activity over the entire Sun}\label{sect:wholesun}

\begin{figure}
    \centering
    \includegraphics[trim=0cm 0cm 0cm 0cm, clip, width=0.5\textwidth]{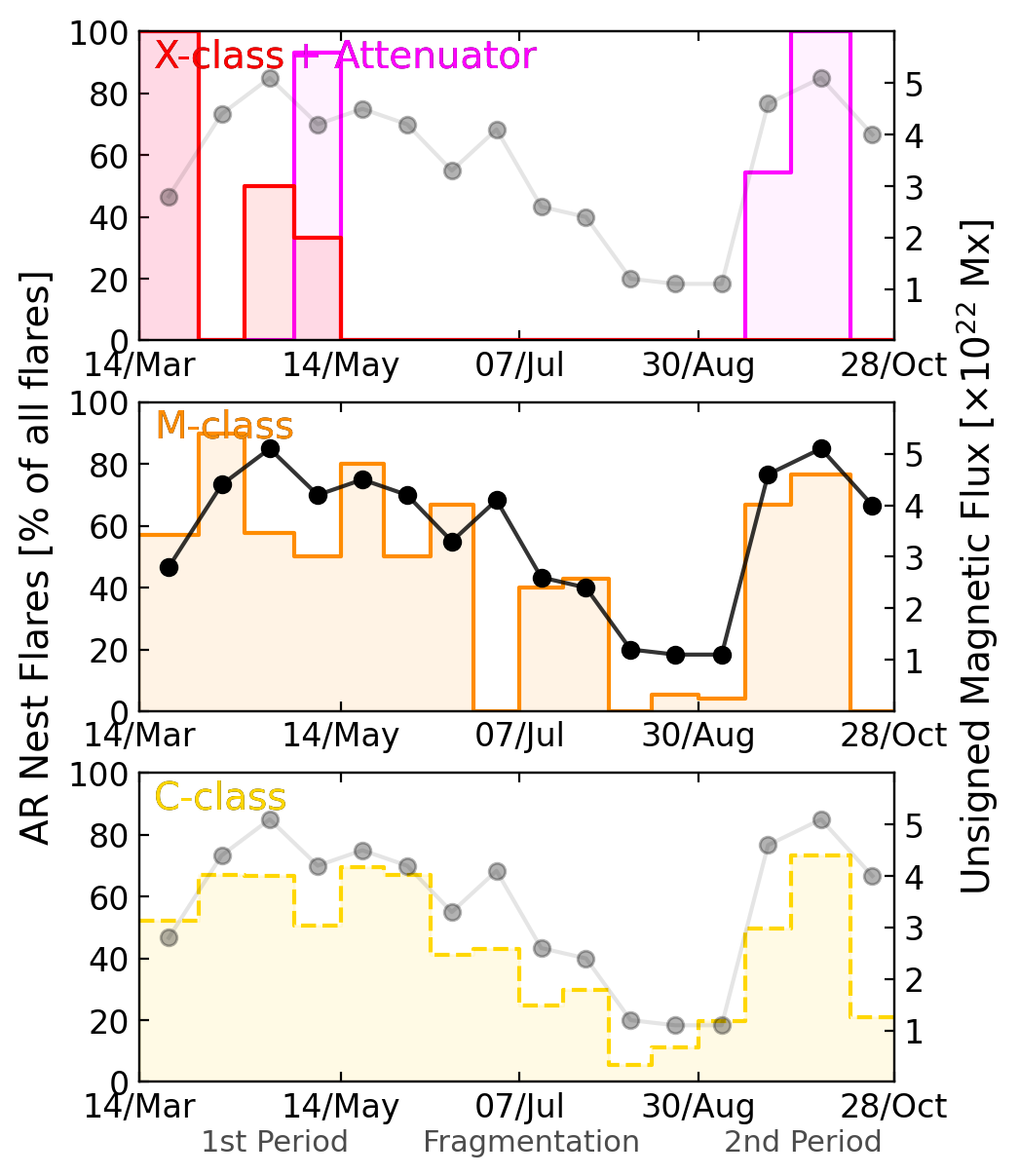}
    \caption{Percentage of solar flares from the AR nest vs. the entire Sun during the window of nearly continuous observations. The unsigned magnetic flux contained within the EUV contour of the AR nest is shown for comparison.}
    \label{fig:totalactivity}
\end{figure}

\begin{figure*}
    \centering
    \includegraphics[trim=0cm 0cm 0cm 0cm, clip, width=\textwidth]{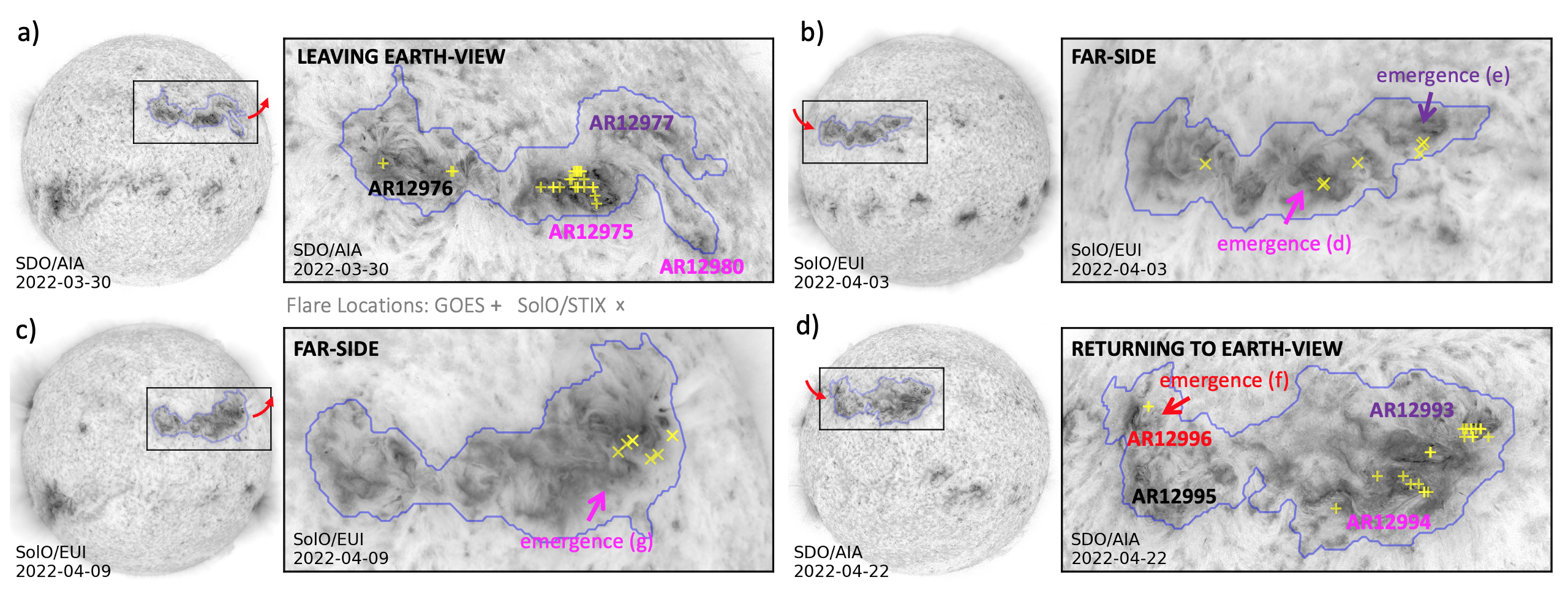}
    \caption{Evolution of individual ARs within the AR nest during the first period of high activity. Panel (a): AR nest rotates from the field of view of Earth towards that of Solar Orbiter. Panels (b) and (c): Evolution of the AR nest whilst on the far side from Earth. Panel (d): AR nest returning to the field of view of Earth. In panels (a) and (d) we show the NOAA AR numbers. The colours indicate a continuity between existing ARs, flux emergence events (labelled to match Fig. \ref{fig:mag_summary}), and the new AR numbers. The solar flares are highlighted with yellow markers, regardless of their GOES flare class.}
    \label{fig:highlight}
\end{figure*}

To highlight the net contribution of the AR nest to the global solar activity, we split the observations from April to October into Carrington maps that combined observations from Solar Orbiter and near-Earth satellites. These Carrington maps required less time to produce than the typical 27.3 days as each observer was approximately 180$^{\circ}$ apart in longitude; a minimum of 14 days to a maximum of 19 days was required for each. Example maps for the first period of high activity are available in Appendix \ref{ap:carrington}. Figure \ref{fig:carr_charts} shows the combined EUV maps from AIA and EUI, along with markers for the solar flares and the EUV contours from Section \ref{sect:identify}. We extracted the area of the EUV contours, computed the unsigned magnetic flux, and summed the number of solar flares originating from the AR nest. These values were then compared to the global unsigned flux and flare frequency. This information is summarised in Table \ref{tab:global_params}. 

The fraction of C-, M-, and X-class (plus attenuated) flares from the AR nest versus the entire Sun during March to October is shown in Fig. \ref{fig:totalactivity}, along with the evolution of the unsigned magnetic flux. From March to May, the AR nest produced 50-70\% of the total C-class flares and 50-90\% of the M-class flares, and it contained 8-16\% of the solar surface magnetic flux. This time period also contained three X-class flares observed by GOES. The other three X-class flares came from two separate ARs elsewhere on the Sun. X-class flares are most commonly associated with high flux emergence rates \citep{schrijver2007characteristic} such as those seen during the first period of high activity. The flare activity decreased across the entire Sun from May to June, and the flare occurrence rate from the AR nest declined to a low in August. Although it did not contain any X-class events, the second high-activity period produced a higher rate of C-class and attenuated flares than the first period. During this time, the AR nest caused 50-70\% of the total C-class flares,  $\sim$70\% of the M-class flares, and the vast majority of flares triggering the STIX attenuator. Averaged over April to October, we found that the AR nest caused $\sim$40\% of all flares over the entire Sun. The AR nest contained 10 of 17 complex flaring regions in 2022 (see Fig. \ref{fig:complexity}) but accounted for only 16.7\% of the available area in the active latitudes (northern and southern).

To show the value of far side observations in this study, we removed the data collected by Solar Orbiter from our analysis and recomputed the number of flares from the AR nest. Using GOES alone, we miss $\sim$ 55\% ($\sim$ 950) of all C-class flares from the AR nest during the near-continuous monitoring window of eight CRs. Each individual CR loses from 31\% to 68\% of C-class flares and from 22\% to 45\% of M-class flares. The most extreme case is CR2255 because the AR nest had a high flaring rate on the far side following the X-class flare on 30 March. The percentage of solar flares that originated in the AR nest versus the total observed flares follows a very similar trend to the combined dataset (it deviates by a few percent). The only exception is CR2255, where the contribution from the AR nest is strongly underestimated (45\% compared to 60\% using the combined dataset) due to the high rate of flaring in the AR nest on the far side.

\subsection{Tracking features during the peak of the flaring activity}\label{sect:highlight}

From the Earth alone, ARs rotate off the visible disk and evolve on the far side for two weeks. Regardless of whether the AR returns after two weeks on the far side, it is given a new NOAA AR number. The lack of continuity makes it difficult to track ARs for long periods of time. In addition, as AR nests are formed of multiple ARs, there is no simple way to associate decaying and emerging regions in the same nest together. The nearly continuous observations presented in this study allowed us to maintain continuity between ARs. To show this, we closely examined one solar rotation in April, during the first period of high activity. 

Figure \ref{fig:highlight} contains four full-disk EUV images: two images from AIA, and another two images from EUI. Panel (a) shows the AR nest leaving the field of view of Earth, and panel (d) shows when it returned. The far side evolution of the AR nest was captured by EUI in panels (b) and (c). When it left the Earth's view in panel (a), the AR nest contained four NOAA ARs, AR12975, AR12976, AR12977, and AR12980. When it was out of view on the far side, panel (b) highlights flux emergence events (d) and (e) in AR12975 and AR12977 (see Fig. \ref{fig:mag_summary}). Then in panel (c), flux emergence event (g) takes place in between AR12975 and AR12980. When the AR nest returned to view in panel (d), due to the flux emergence events on the far side, 1) AR12977 ($\beta$) had become AR12993 ($\beta\gamma$), 2) AR12975 ($\beta\gamma\delta$) merged with AR12980 to become AR12994 ($\beta\gamma$), and 3) AR12976 ($\beta\delta$) decayed slightly to form AR12995 ($\beta$). Whilst the nest rotated into view, an additional flux emergence event (f) created AR12996 ($\beta$) just above AR12995.

The change in the free magnetic energy due to the flux emergence events is emphasised by the evolving solar flare sources. Panel (a) shows that solar flares were clustered around AR12975 ($\beta\gamma\delta$ configuration), whilst panel (d) shows the flares that occurred amongst the two $\beta\gamma$ ARs that had experienced flux emergence on the far side. This type of dataset will in the future allow us to study how magnetic complexity can be built within AR nests. This has consequences for the free magnetic energy that is available to power solar flares \citep[see][]{shibata1998evidence}. This example shows that we can maintain continuity between ARs despite the intermediary flux emergence events. A deeper exploration of this dataset is warranted and might motivate new AR classifications that take the history and surroundings of a region into account.

\subsection{Antipodal activity}\label{sect:antipodal}

Previous studies have observed that AR nests form anti-symmetrically across the northern and southern hemispheres \citep{bai1987distribution,bai1990solar,pojoga2002clustering}, with a 180$^{\circ}$ phase shift in Carrington longitude. In Fig. \ref{fig:average_sum} we also identified hot spots of activity in the southern hemisphere. These two regions were far away from the antipodal longitude of 240$^{\circ}$ Carrington longitude, about 60$^{\circ}$ on either side. However, when we focus on the larger M- and X-class flares in the southern hemisphere, we find that they were scattered much closer to the antipodal longitude. The X-class flares in the southern hemisphere were located almost exactly at 240$^{\circ}$ Carrington longitude. Activity continued in this region during 2023 (visible in Fig. \ref{fig:beforeafter}, panel c). As activity in this region was more pronounced in the second half of 2022, it is possible that the averaging window in Fig. \ref{fig:average_sum} was not sufficient to capture the southern AR nest. Perhaps it was in the process of flip-flopping from the northern to southern hemisphere \citep[e.g.][]{berdyugina2005spot}. Further investigation is needed to ascertain whether our northern AR nest had an antipodal southern AR nest.

\section{Conclusions}

AR nests strongly influence the spatial distribution of solar flares. Repeated flux emergence events fuel flaring activity over several solar rotations. AR nests appear to form in a coherent manner, which points towards the underlying global solar dynamo \citep{brun2017magnetism}. We identified an AR nest in 2022 that was located in the northern solar hemisphere around 90$^{\circ}$ Carrington longitude. We combined observations from ESA's Solar Orbiter with near-Earth observatories to monitor the magnetic activity of the AR nest. From April to October, Solar Orbiter was located on the far- side of the Sun to Earth, which facilitated nearly continuous observations (see Fig. \ref{fig:soloOrbit}). This is similar to the STEREO era, when the twin A and B spacecraft were spread out along Earth's orbit to produce a 360$^{\circ}$ view of the Sun in February 2011. However, the STEREO spacecraft were unable to characterise solar flares or measure the photospheric magnetic field. With Solar Orbiter, we were able to monitor changes in the EUV morphology, photospheric magnetic field, and solar flare properties. 

The AR nest underwent multiple flux emergence events, some of which occurred while the nest was on the far side to Earth and were captured by Solar Orbiter. This may explain why the region contained the majority of complex flaring NOAA ARs in 2022. At the end of April, the AR nest reached its highest unsigned flux of $\sim 5\times 10^{22}$Mx, contained two complex ARs (one $\beta\gamma\delta$ and one $\beta\gamma$), and produced large solar flares at a rate of around 15 M-class flares and one X-class flare per week. The AR nest also significantly affected the coronal magnetic field, as shown in Fig. \ref{fig:HCS_evol} and further investigated by \cite{finley2024nested}. It essentially anchored the heliospheric current sheet above it. 

By combining the solar flares detected at Earth with those on the far side from Solar Orbiter, we were able to measure the majority of flares on the Sun from April to October. By comparing the number of flares originating in the AR nest to the rest of the solar surface, we showed that the AR nest produced 40-70\% of all solar flares from March to June as well as from September to October but contained only about 8-16\% of the total unsigned photospheric magnetic flux (see Table \ref{tab:global_params}). There were additional bursts of solar activity in February and December from the AR nest, but we lacked far side observations during these times. Averaged over 2022, the AR nest produced 40\% of observed solar flares. The longitudinal extent, interconnected magnetic field, and high rate of flaring in AR nests might also influence the likelihood of sympathetic flaring. Previous work found a peak in the longitudinal separation of sympathetic flares around $30^{\circ}$ \citep{guite2025flaring}, which is comparable with the extent of the AR nest in this study. 

Whilst searching for activity in the nesting region before and after 2022, as 2022 was chosen to focus on the nearly continuous coverage facilitated by Solar Orbiter, we found flux emergence events in 2020, 2021, and 2023 (see Fig. \ref{fig:beforeafter}). Whether these emergences are connected to the AR nest in 2022 is unclear, but this might support previous studies that suggested that AR nests are coherent over much longer timescales. More work is needed to build statistics on the formation and lifetimes of AR nests, taking advantage of observations from the STEREO era \citep{kaiser2008stereo} and the more recent observations from Solar Orbiter. Continued joint-observation studies are required to assess the flux emergence and solar flare occurrence rates. It remains to be definitively proven whether AR nests produce more complex ($\beta\gamma\delta$) ARs than isolated flux emergence events, although this was the case for the AR nest in 2022. Future works should also investigate the occurrence rates and properties of coronal mass ejections that emerge from AR nests versus isolated ARs, as they represent a severe space weather risk.

\begin{acknowledgements}
The authors thank Säm Krucker, Nicole Vilmer, and Olivier Limousin for continuous discussion on STIX data analysis, the inner workings of the instrument and the physics of solar flares. We thank the STIX team for developing the analysis tools for STIX and maintaining the Data Center. 
We thank Laura Hayes, Hannah Collier and Andrea Battaglia for making available the tools to generate the STIX flare list.
This research has received funding from the European Research Council (ERC) under the European Union’s Horizon 2020 research and innovation programme (grant agreement No 810218 WHOLESUN), in addition to funding by the Centre National d'Etudes Spatiales (CNES) Solar Orbiter, the French Agence Nationale de la Recherche (ANR) project STORMGENESIS \#ANR-22-CE31-0013-01, and the Institut National des Sciences de l'Univers (INSU) via the Action Thématique Soleil-Terre (ATST).
Data supplied courtesy of the SDO/HMI and SDO/AIA consortia. SDO is the first mission to be launched for NASA's Living With a Star (LWS) Program.
Data manipulation was performed using the numpy \citep{2020NumPy-Array}, scipy \citep{2020SciPy-NMeth}, and pySHTOOLS \citep{wieczorek2018shtools} python packages.
Figures in this work are produced using the python package matplotlib \citep{hunter2007matplotlib}.
\end{acknowledgements}

%
%

\bibliographystyle{aa}
\bibliography{adam}

\onecolumn
\begin{appendix}

\section{Stack plots of solar activity in 2022} \label{ap:stack_plots}

\begin{figure*}[h!]
    \centering
    \includegraphics[trim=0cm 0cm 0cm 0cm, clip, width=\textwidth]{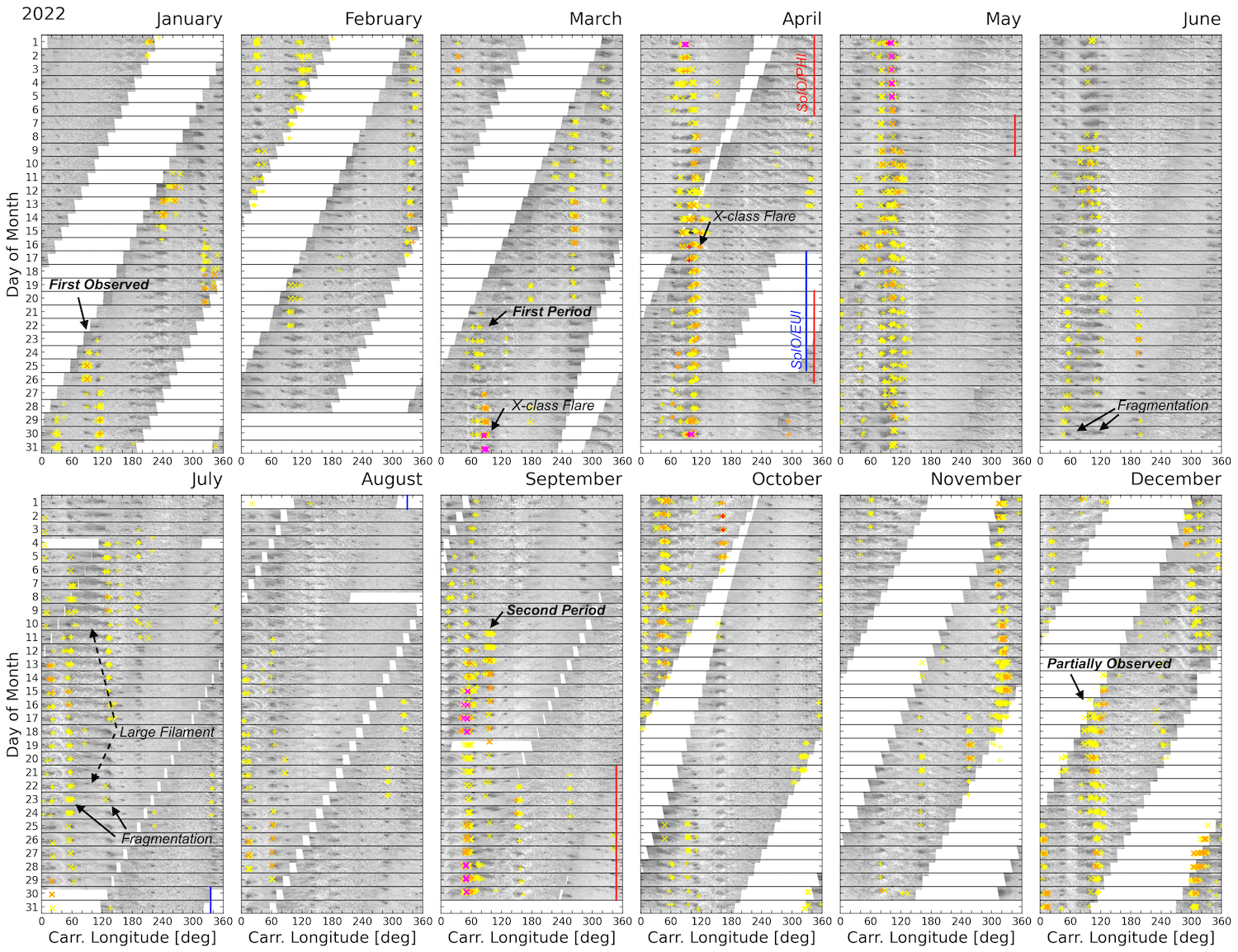}
    \caption{Monthly stack-plots of solar activity in the northern hemisphere. EUV 304$\AA$ observations using the daily synoptic observations from AIA and EUI are shown in greyscale for context. The location of solar flares observed by GOES and STIX are plotted with coloured markers. The colour corresponds to the GOES flare class of X, M, and C, in red, orange, and yellow, respectively. Magenta markers highlight STIX flares that triggered the attentuator. Red and blue vertical bars highlight observing gaps larger than two days for PHI and EUI, respectively.}
    \label{fig:flares_stacked}
\end{figure*}

\begin{figure*}[h!]
    \centering
    \includegraphics[trim=0cm 0cm 0cm 0cm, clip, width=\textwidth]{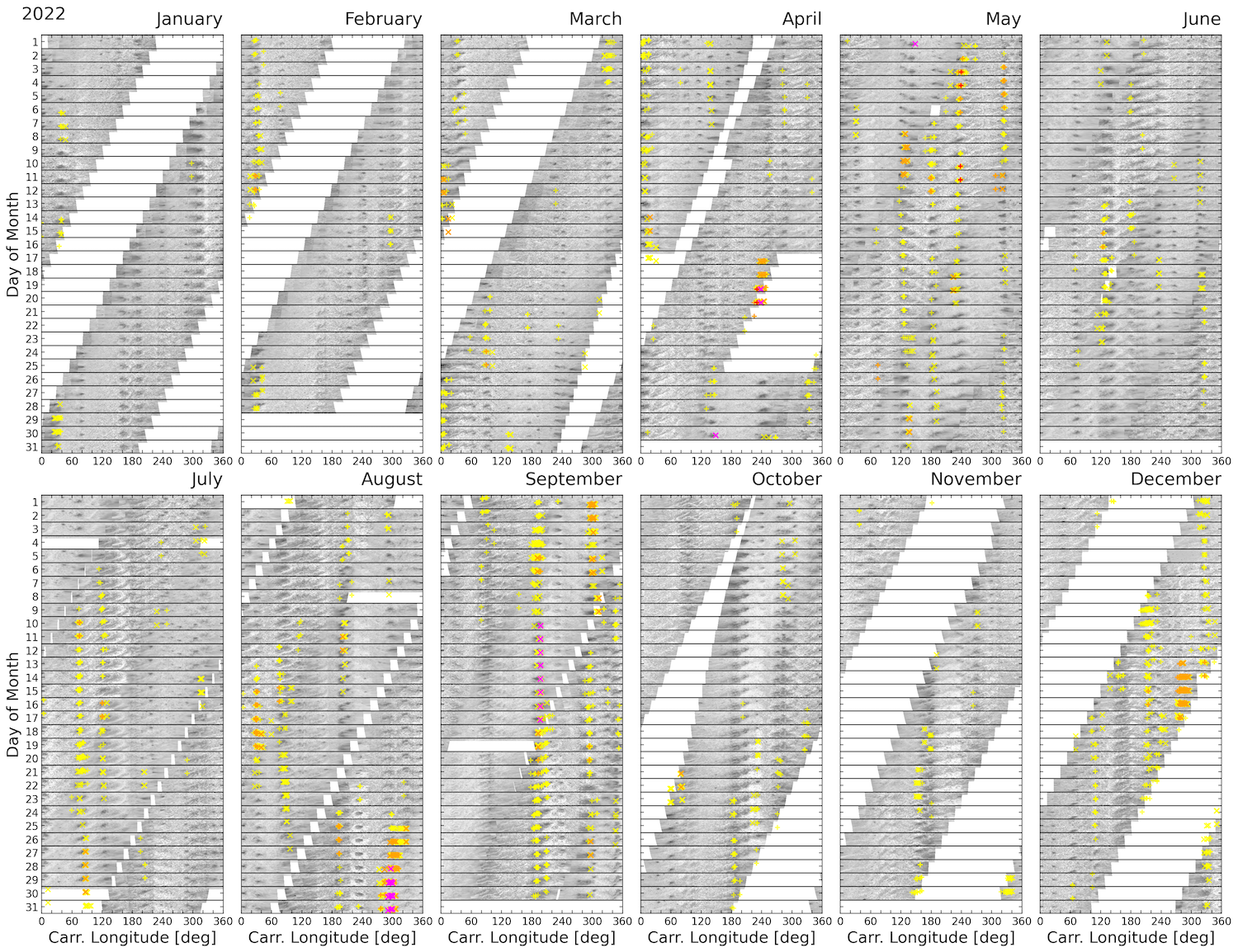}
    \caption{Same as Fig. \ref{fig:flares_stacked} but for the southern hemisphere.}
    \label{fig:flares_stacked_south}
\end{figure*}

EUV observations from AIA and EUI were combined for each day in 2022, when available, in addition to combining the flare lists from GOES and STIX. This information is displayed in time-longitude stack plots in Figs. \ref{fig:flares_stacked} and \ref{fig:flares_stacked_south} for the northern and southern active latitudes respectively. This allows the available observations and data gaps to be visualised. Gaps in EUI and PHI that were longer than two days are marked for reference. Flaring activity is clustered in the northern hemisphere around 60-90$^{\circ}$ Carrington longitude. However, the flare occurrence rate in this region decreases in February, August and November. In the southern hemisphere, flaring activity is visibly less structured with the strongest flares sources at either 180$^{\circ}$ or 300$^{\circ}$. Some X-class flares were produced near 240$^{\circ}$, which is antipodal to the clustering of activity in the northern hemisphere at 90$^{\circ}$.

\clearpage

\section{Encounter 11 of Parker Solar Probe}\label{ap:psp_e11}

\begin{figure*}[h!]
    \centering
    \includegraphics[trim=0cm 0cm 0cm 0cm, clip, width=\textwidth]{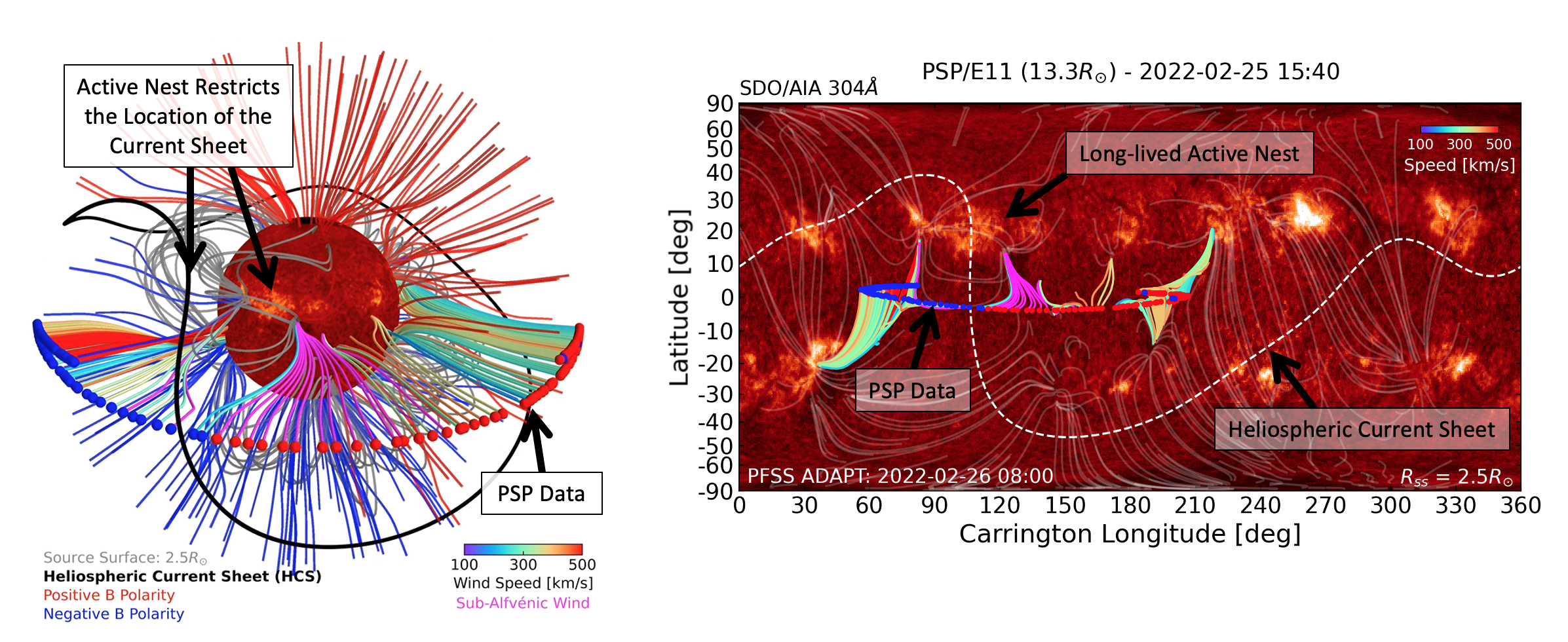}
    \caption{Potential field source surface extrapolation of the coronal magnetic field during encounter 11 of NASA's Parker Solar Probe. The observed magnetic field polarity is mapped to the source surface for comparison with the heliospheric current sheet. The magnetic field lines that trace down to the source regions are coloured by the observed solar wind speed.}
    \label{fig:psp_crossing}
\end{figure*}

During 2022, the AR nest played a significant role in sculpting the solar coronal magnetic field, as shown in Fig. \ref{fig:HCS_evol}. The strong opposite polarity magnetic fields that were maintained within the AR nest anchored the heliospheric current sheet in place for several solar rotations \citep[discussed in][]{finley2024nested}. NASA's Parker Solar Probe passed through the inclined current sheet above the AR nest in late February. Figure \ref{fig:psp_crossing} shows the in-situ solar wind speed and magnetic field polarity measured by Parker Solar Probe during encounter 11, mapped down to the solar surface using ballistic back-projection and potential field source surface mapping. The reversal of the magnetic field polarity directly above the AR nest is clearly identified. Around the same time, Solar Orbiter also scanned across this region facilitating coordinated observations between the two in-situ observers; used to explore the heating and acceleration of the solar wind \citep[see][]{rivera2024situ}. The stability of the coronal magnetic field around this region increased the likelihood of finding a common source region. This highlights the importance of monitoring AR nests in order to take advantage of favourable magnetic connectivity between in-situ and remote-sensing observers.

\clearpage

\section{Activity in 2020, 2021, and 2023} \label{ap:beforeafter}

\begin{figure*}[h!]
    \centering
    \includegraphics[trim=0cm 0cm 0cm 0cm, clip, width=0.95\textwidth]{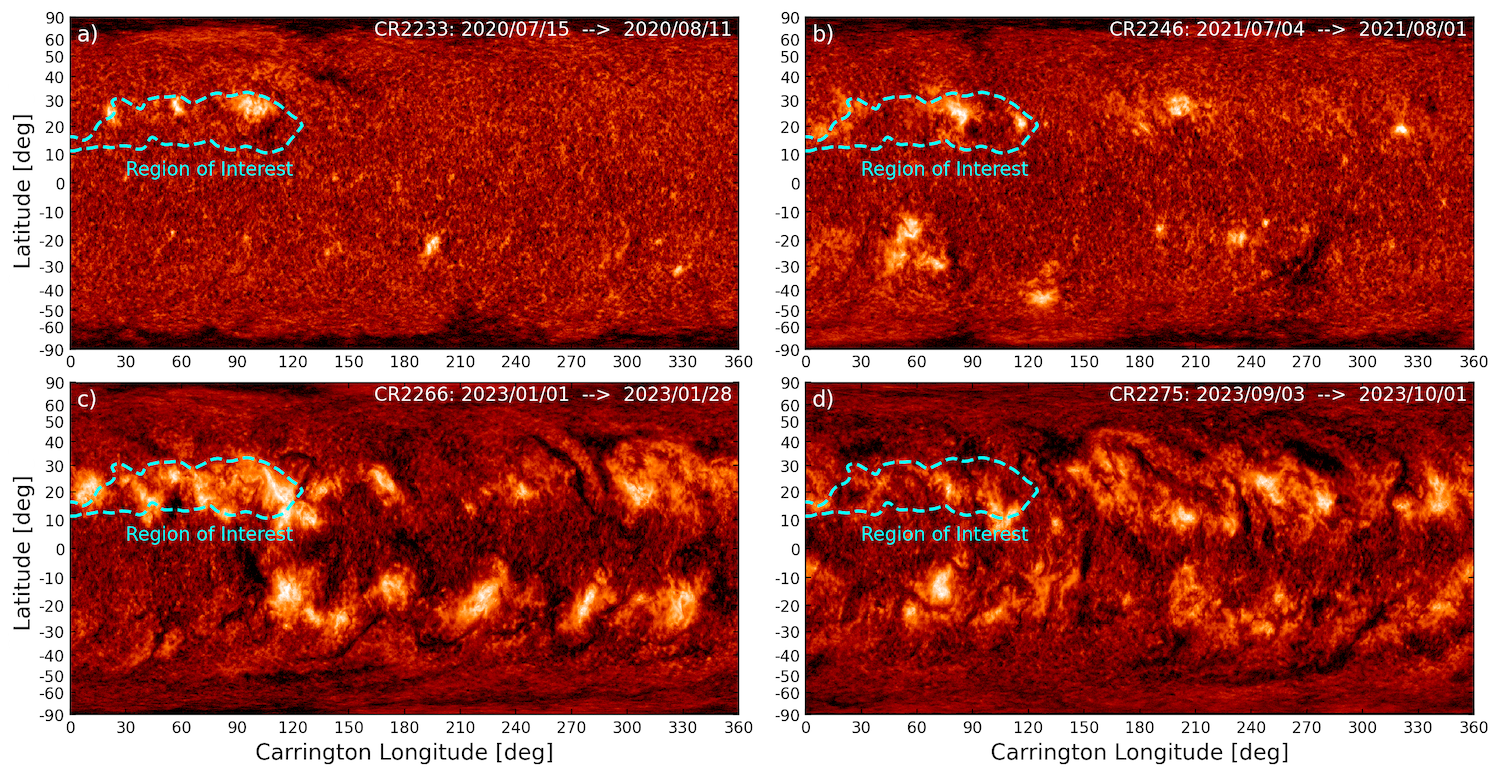}
    \caption{Carrington maps of EUV 304$\AA$ from AIA highlighting the presence of magnetic activity in the region of interest before and after 2022. Panels (a) and (b) correspond to July-August 2020 and July 2021, respectively. Panels (c) and (d) show activity from the AR nest in 2022 persisting into 2023. The AR nest identified in Fig. \ref{fig:average_sum} is shown with a dashed cyan contour.}
    \label{fig:beforeafter}
\end{figure*}

The choice to study 2022 was made to take advantage of the far side observations from Solar Orbiter, however this was arbitrary. If we expand our time range to encompass the rising phase of solar cycle 25 (from 2020 to 2023) then we find additional flux emergence events in the region of interest identified in 2022 (contour from Fig. \ref{fig:average_sum}). Figure \ref{fig:beforeafter} shows some examples of ARs that formed in this area. The first significant activity of solar cycle 25 in the northern hemisphere appeared in this region, panel (a). More activity was observed in 2021, panel (b). After 2022, the entire region of interest became active in January 2023, panel (c), and contained some flux emergence in September 2023, panel (d). From 2020 to 2023, the latitude of AR emergence in this area evolved following the solar butterfly pattern from around $25^{\circ}$ to $10^{\circ}$. A detailed analysis of these observations is left for future work, however the continuity in activity may suggest that AR nests could in future be forecast.

\section{Carrington maps from March to July} \label{ap:carrington}

\begin{figure*}[h!]
    \centering
    \includegraphics[trim=0cm 0cm 0cm 0cm, clip, width=\textwidth]{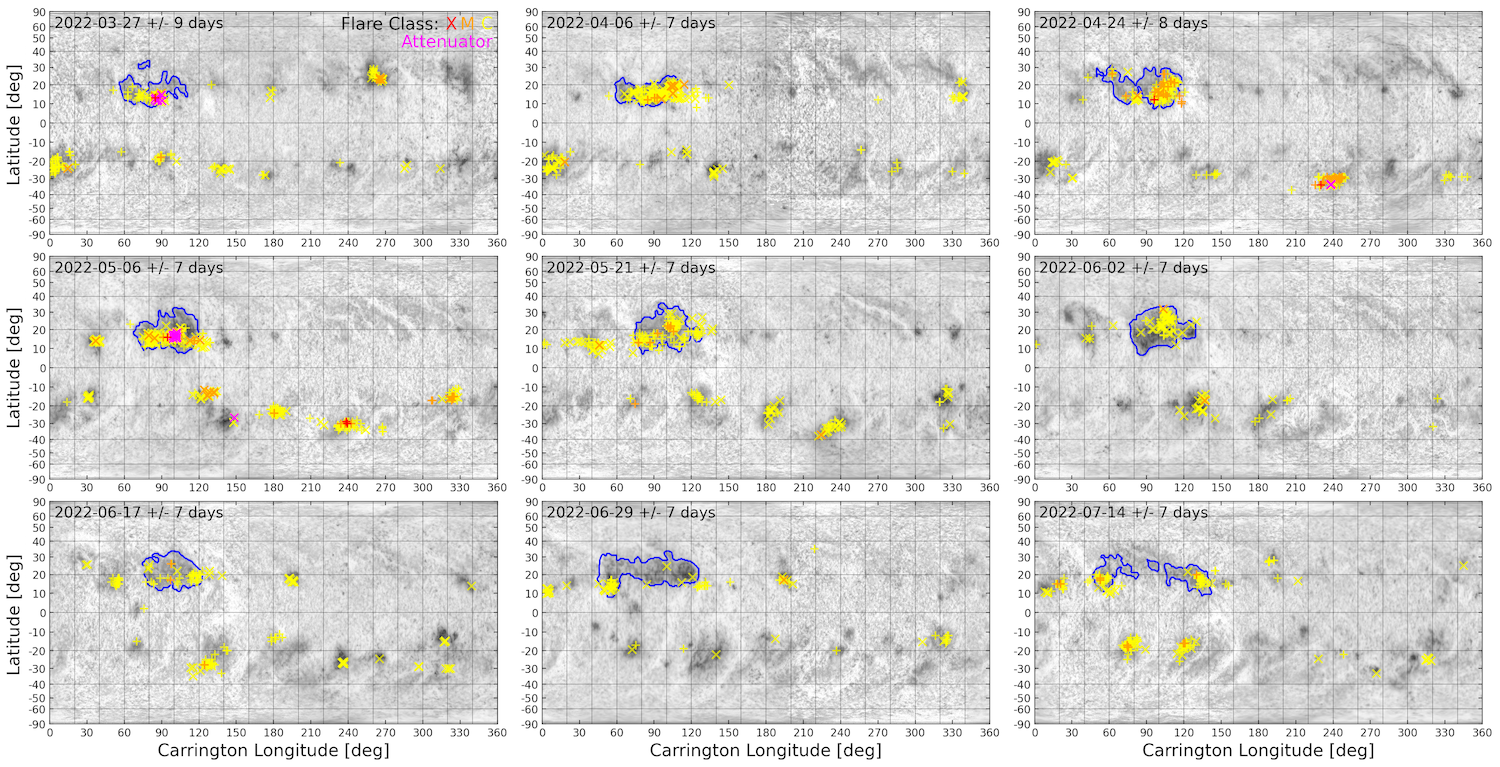}
    \caption{Combined EUV 304$\AA$ Carrington maps (grey scale) with the location of solar flares observed by GOES and STIX marked with crossed that are coloured by flare class. The AR nest is outlined with a blue contour. This series of maps covers the first period of high activity from the end of March until the fragmentation and filament formation in July.}
    \label{fig:carr_charts}
\end{figure*}

We split the period of near-continuous observations into roughly two week intervals, the time taken to build up a Carrington map of the entire solar surface in EUV with AIA and EUI. The intervals were chosen such that the AR nest passes the central meridian of one of the observers for each scan, i.e., the AR nest is mapped by AIA in the first map, then EUI in the next, etc. The combined maps are shown in Fig. \ref{fig:carr_charts}, these correspond to the same time intervals as in Table \ref{tab:global_params}. In each Carrington map the EUV contour of the AR nest from Section \ref{sect:identify} is over-plotted along with the location of flares observed by GOES and STIX. From each interval, the number of flares being produced by the AR nest is evaluated against the total (global) value. During this time period, the AR nest caused 50-70\% of all flares on the Sun.

\end{appendix}
\end{document}